\documentclass[preprint2,iop,numberedappendix,twocolappendix,appendixfloats]{emulateapj}

\usepackage[caption=false]{subfig}
\usepackage{amsmath}
\usepackage{footnote}
\bibpunct{(}{)}{;}{a}{}{,} 
\newcommand{\dif}{\mathrm{d}}

\shorttitle{Foregrounds in EoR Power Spectra}
\shortauthors{Thyagarajan et~al.}

\def\ASU{\altaffilmark{1}}
\def\ASUtxt{\altaffiltext{1}{Arizona State University, School of Earth and Space Exploration, Tempe, AZ 85287, USA}}

\def\myemail{\altaffilmark{*}}
\def\myemailtxt{\altaffiltext{*}{e-mail: t\_nithyanandan@asu.edu}}

\def\UW{\altaffilmark{2}}
\def\UWtxt{\altaffiltext{2}{University of Washington, Department of Physics, Seattle, WA 98195, USA}}

\def\SKASA{\altaffilmark{3}}
\def\SKASAtxt{\altaffiltext{3}{Square Kilometre Array South Africa (SKA SA), Park Road, Pinelands 7405, South Africa}}

\def\RU{\altaffilmark{4}}
\def\RUtxt{\altaffiltext{4}{Department of Physics and Electronics, Rhodes University, Grahamstown 6140, South Africa}}

\def\CfA{\altaffilmark{5}}
\def\CfAtxt{\altaffiltext{5}{Harvard-Smithsonian Center for Astrophysics, Cambridge, MA 02138, USA}}

\def\ANU{\altaffilmark{6}}
\def\ANUtxt{\altaffiltext{6}{Australian National University, Research School of Astronomy and Astrophysics, Canberra, ACT 2611, Australia}}

\def\CAASTRO{\altaffilmark{7}}
\def\CAASTROtxt{\altaffiltext{7}{ARC Centre of Excellence for All-sky Astrophysics (CAASTRO)}}

\def\Haystack{\altaffilmark{8}}
\def\Haystacktxt{\altaffiltext{8}{MIT Haystack Observatory, Westford, MA 01886, USA}}

\def\MIT{\altaffilmark{9}}
\def\MITtxt{\altaffiltext{9}{MIT Kavli Institute for Astrophysics and Space Research, Cambridge, MA 02139, USA}}

\def\Curtin{\altaffilmark{10}}
\def\Curtintxt{\altaffiltext{10}{International Centre for Radio Astronomy Research, Curtin University, Perth, WA 6845, Australia}}

\def\Victoria{\altaffilmark{11}}
\def\Victoriatxt{\altaffiltext{11}{Victoria University of Wellington, School of Chemical \& Physical Sciences, Wellington 6140, New Zealand}}

\def\UWisc{\altaffilmark{12}}
\def\UWisctxt{\altaffiltext{12}{University of Wisconsin--Milwaukee, Department of Physics, Milwaukee, WI 53201, USA}}

\def\UMichigan{\altaffilmark{13}}
\def\UMichigantxt{\altaffiltext{13}{University of Michigan, Department of Atmospheric, Oceanic and Space Sciences, Ann Arbor, MI 48109, USA}}

\def\UMelbourne{\altaffilmark{14}}
\def\UMelbournetxt{\altaffiltext{14}{The University of Melbourne, School of Physics, Parkville, VIC 3010, Australia}}

\def\USydney{\altaffilmark{15}}
\def\USydneytxt{\altaffiltext{15}{The University of Sydney, Sydney Institute for Astronomy, School of Physics, NSW 2006, Australia}}

\def\CASS{\altaffilmark{16}}
\def\CASStxt{\altaffiltext{16}{CSIRO Astronomy and Space Science (CASS), PO Box 76, Epping, NSW 1710, Australia}}

\def\Tata{\altaffilmark{17}}
\def\Tatatxt{\altaffiltext{17}{National Centre for Radio Astrophysics, Tata Institute for Fundamental Research, Pune 411007, India}}

\def\RRI{\altaffilmark{18}}
\def\RRItxt{\altaffiltext{18}{Raman Research Institute, Bangalore 560080, India}}

\def\NRAO{\altaffilmark{19}}
\def\NRAOtxt{\altaffiltext{19}{National Radio Astronomy Observatory, Charlottesville and Greenbank, USA}}

\def\UWA{\altaffilmark{20}}
\def\UWAtxt{\altaffiltext{20}{International Centre for Radio Astronomy Research, University of Western Australia, Crawley, WA 6009, Australia}}

\begin{document}

\title{Foregrounds in Wide-Field Redshifted 21~cm Power Spectra}


\author{
Nithyanandan~Thyagarajan\ASU\myemail,
Daniel~C.~Jacobs\ASU,
Judd~D.~Bowman\ASU,
N.~Barry\UW,
A.~P.~Beardsley\UW,
G.~Bernardi\SKASA$^,$\RU$^,$\CfA,
F.~Briggs\ANU$^,$\CAASTRO,
R.~J.~Cappallo\Haystack, 
P.~Carroll\UW,
B.~E.~Corey\Haystack, 
A.~de~Oliveira-Costa\MIT,
Joshua~S.~Dillon\MIT,
D.~Emrich\Curtin,
A.~Ewall-Wice\MIT,
L.~Feng\MIT,
R.~Goeke\MIT,
L.~J.~Greenhill\CfA,
B.~J.~Hazelton\UW, 
J.~N.~Hewitt\MIT,
N.~Hurley-Walker\Curtin,
M.~Johnston-Hollitt\Victoria,
D.~L.~Kaplan\UWisc, 
J.~C.~Kasper\UMichigan$^,$\CfA, 
Han-Seek Kim\UMelbourne$^,$\CAASTRO,
P.~Kittiwisit\ASU,
E.~Kratzenberg\Haystack, 
E.~Lenc\USydney$^,$\CAASTRO,
J.~Line\UMelbourne$^,$\CAASTRO,
A.~Loeb\CfA,
C.~J.~Lonsdale\Haystack, 
M.~J.~Lynch\Curtin, 
B.~McKinley\UMelbourne$^,$\CAASTRO,
S.~R.~McWhirter\Haystack,
D.~A.~Mitchell\CASS$^,$\CAASTRO, 
M.~F.~Morales\UW, 
E.~Morgan\MIT, 
A.~R.~Neben\MIT,
D.~Oberoi\Tata, 
A.~R.~Offringa\ANU$^,$\CAASTRO, 
S.~M.~Ord\Curtin$^,$\CAASTRO,
Sourabh Paul\RRI,
B.~Pindor\UMelbourne$^,$\CAASTRO,
J.~C.~Pober\UW,
T.~Prabu\RRI, 
P.~Procopio\UMelbourne$^,$\CAASTRO,
J.~Riding\UMelbourne$^,$\CAASTRO,
A.~E.~E.~Rogers\Haystack, 
A.~Roshi\NRAO, 
N.~Udaya~Shankar\RRI, 
Shiv~K.~Sethi\RRI,
K.~S.~Srivani\RRI, 
R.~Subrahmanyan\RRI$^,$\CAASTRO, 
I.~S.~Sullivan\UW,
M.~Tegmark\MIT,
S.~J.~Tingay\Curtin$^,$\CAASTRO, 
C.~M.~Trott\Curtin$^,$\CAASTRO,
M.~Waterson\Curtin$^,$\ANU,
R.~B.~Wayth\Curtin$^,$\CAASTRO, 
R.~L.~Webster\UMelbourne$^,$\CAASTRO, 
A.~R.~Whitney\Haystack, 
A.~Williams\Curtin, 
C.~L.~Williams\MIT,
C.~Wu\UWA,
J.~S.~B.~Wyithe\UMelbourne$^,$\CAASTRO
}

\ASUtxt
\UWtxt
\SKASAtxt
\RUtxt
\CfAtxt
\ANUtxt
\CAASTROtxt
\Haystacktxt
\MITtxt
\Curtintxt
\Victoriatxt
\UWisctxt
\UMichigantxt
\UMelbournetxt
\USydneytxt
\CASStxt
\Tatatxt
\RRItxt
\NRAOtxt
\UWAtxt
\myemailtxt



\begin{abstract}

Detection of 21~cm emission of H~{\sc i} from the epoch of reionization, at redshifts $z>6$, is limited primarily by foreground emission. We investigate the signatures of wide-field measurements and an all-sky foreground model using the delay spectrum technique that maps the measurements to foreground object locations through signal delays between antenna pairs. We demonstrate interferometric measurements are inherently sensitive to all scales, including the largest angular scales, owing to the nature of wide-field measurements. These wide-field effects are generic to all observations but antenna shapes impact their amplitudes substantially. A dish-shaped antenna yields the most desirable features from a foreground contamination viewpoint, relative to a dipole or a phased array. Comparing data from recent Murchison Widefield Array observations, we demonstrate that the foreground signatures that have the largest impact on the H~{\sc i} signal arise from power received far away from the primary field of view. We identify diffuse emission near the horizon as a significant contributing factor, even on wide antenna spacings that usually represent structures on small scales. For signals entering through the primary field of view, compact emission dominates the foreground contamination. These two mechanisms imprint a characteristic {\it pitchfork} signature on the ``foreground wedge'' in Fourier delay space. Based on these results, we propose that selective down-weighting of data based on antenna spacing and time can mitigate foreground contamination substantially by a factor $\sim 100$ with negligible loss of sensitivity. 

\end{abstract}
 

\section{Introduction}\label{intro}

At the end of the recombination epoch, the Universe was completely neutral. This period, referred to as the {\it Dark Ages} in the Universe's history, is characterized by the localized accumulation of matter under the influence of gravity. And it ended with the formation of the first stars and galaxies which started emitting ultra-violet and X-ray radiation, thereby reionizing the neutral medium in their surroundings. This commenced the epoch of reionization (EoR) -- a period of nonlinear growth of matter density perturbations and astrophysical evolution. 

Observing redshifted 21~cm radiation generated by the spin flip transition of H~{\sc i} has been identified as a direct probe of the EoR \citep{sun72,sco90,mad97,toz00,ili02}. Detecting this signal has recently emerged as a very promising experiment to fill the gaps in our understanding of the Universe's history.  

Sensitive instruments such as the Square Kilometre Array (SKA) are required for direct observation and tomography of redshifted H~{\sc i}. Numerous pathfinders and precursors to the SKA such as the Murchison Widefield Array \citep[MWA;][]{lon09,tin13,bow13}, the Low Frequency Array \citep[LOFAR;][]{van13}, and the Precision Array for Probing the Epoch of Reionization \citep[PAPER;][]{par10} have become operational with enough sensitivity for a statistical detection of the EoR H~{\sc i} power spectrum \citep{bow06,par12a,bea13,dil13,thy13,pob14}. The Hydrogen Epoch of Reionization Array\footnote{\url{http://reionization.org/}} (HERA) is currently under construction using new insights gained with the MWA and PAPER.

A key challenge in the statistical detection of the redshifted H~{\sc i} 21~cm signal, via the spatial power spectrum of temperature fluctuations, arises from the contamination by Galactic and extragalactic foregrounds \citep[see, e.g.,][]{dim02,zal04,fur06,ali08,ber09,ber10,gho12}. \citet{mor04} show that the inherent isotropy and symmetry of the EoR signal in frequency and spatial wavenumber ($k$) space make it distinguishable from sources of contamination which are isolated to certain $k$ modes by virtue of their inherent spectral smoothness \citep{mor06,bow09,liu11,par12b,dil13,pob13}. Since this contamination is expected to be several orders of magnitude stronger than the underlying EoR H~{\sc i} signal, it is critical to characterize foregrounds precisely in order to reduce their impact on EoR H~{\sc i} power spectrum detection sensitivity. 

Considerable effort is being made toward understanding the $k$-space behavior of foreground signatures in the observed power spectrum and formulating robust estimators of the true power spectrum \citep{bow09,liu09,dat10,liu11,mor12,tro12,pob13,thy13,dil14,liu14a,liu14b}. A model that provides a generic explanation for the observed foreground power spectrum has emerged, whereby the wide-field (and chromatic) response of the instrument causes the power in smooth spectrum foregrounds to occupy higher $k$-modes into the so-called ``wedge''. The conservative foreground strategy, referred to as {\it avoidance}, that has developed alongside this work is to discard $k$-modes which could be contaminated \citep[e.g.,][]{par14}. The more aggressive alternative is to subtract a sky model and regain access to modes that would be discarded by {\it avoidance}. In both cases, which parts of the sky are most critical to either {\it avoid} or {\it subtract} has remained largely uncertain. Here, we focus primarily on extending the {\it avoidance} strategy by identifying foreground components at greatest risk to ``leak'' from foreground modes to EoR modes and proposing a scheme for down-weighting these components.

Foregrounds with intrinsic deviations from spectral smoothness, instruments with high chromaticity, polarization leakage, calibration errors, or approximations in power spectrum analyses can contaminate the true EoR H~{\sc i} power spectrum. Here we use existing catalogs and a high fidelity instrumental model to capture both foreground and instrumental chromaticity. To decouple these effects from possible analysis effects, such as those pointed out by \citet{haz13}, we compute power spectra using a {\it per-baseline} approach of \citet{par12b}. This approximates the power spectrum as the inverse Fourier transform of the spectra generated by the instrument's correlator. 

In \S\ref{sec:delay-spectrum} we provide an overview of the delay spectrum technique. We investigate signatures generic to all wide-field measurements of EoR power spectra in \S\ref{sec:wide-field}. In \S\ref{sec:sim}, we present the foreground model and a variety of instrument models to rank antenna shapes based on foreground contamination. In \S\ref{sec:MWA}, we describe the MWA setup, summarize the observing parameters, and present the resulting data. Simulations using these observing parameters are compared with the data and analyzed for foreground signatures. We report two important findings: foregrounds that most severely obscure the redshifted 21~cm power spectrum are not caused by emission in the central field of view, but rather by bright objects from near the horizon; and, diffuse Galactic emission plays a significant role hitherto unpredicted. In \S\ref{sec:fg-grading}, we offer an initial description of a more precise foreground avoidance technique that minimizes foreground contamination using prior knowledge of the sky to down-weight adversely contaminated baselines. We present a summary of our work and findings in \S\ref{sec:summary}.

\section{Delay Spectrum}\label{sec:delay-spectrum}

We provide a short overview of the delay spectrum technique \citep{par12a,par12b}. 

Interferometer array data known as {\it visibilities}, $V_b(f)$, represent correlations between time-series of electric fields measured by different antenna pairs with separation vectors $\boldsymbol{b}$ and then Fourier transformed along the time axis to obtain a spectrum along the frequency ($f$) axis. If $I(\hat{\boldsymbol{s}},f)$ and $A(\hat{\boldsymbol{s}},f)$ are the sky brightness and antenna's directional power pattern, respectively, at different frequencies as a function of direction on the sky denoted by the unit vector ($\hat{\boldsymbol{s}}$), and $W_\textrm{i}(f)$ denotes instrumental bandpass weights, then $V_b(f)$ can be written as:
\begin{align}\label{eqn:obsvis}
  V_b(f) &= \iint\limits_\textrm{sky} A(\hat{\boldsymbol{s}},f)\,I(\hat{\boldsymbol{s}},f)\,W_\textrm{i}(f)\,e^{-i2\pi f\frac{\boldsymbol{b}\cdot\hat{\boldsymbol{s}}}{c}}\,\dif\Omega,
\end{align}
where, $c$ is the speed of light, and $\dif\Omega$ is the solid angle element to which $\hat{\boldsymbol{s}}$ is the unit normal vector. This equation is valid in general, including wide-field measurements, and is a slight adaptation from \citet{van34}, \citet{zer38}, and \citet{tho01}.

The {\it delay spectrum}, $\tilde{V}_b(\tau)$, is defined as the inverse Fourier transform of $V_b(f)$ along the frequency coordinate:
\begin{align}\label{eqn:delay-transform}
  \tilde{V}_b(\tau) &\equiv \int V_b(f)\,W(f)\,e^{i2\pi f\tau}\,\dif f,
\end{align}
where, $W(f)$ is a spectral weighting function which can be chosen to control the quality of the delay spectrum \citep{ved12,thy13}, and $\tau$ represents the signal delay between antenna pairs:
\begin{equation}\label{eqn:delay}
  \tau = \frac{\boldsymbol{b}\cdot\hat{\boldsymbol{s}}}{c}.
\end{equation}
The delay transform conventions used in this paper are described in appendix~\ref{sec:delay-conventions}. $\tilde{V}_b(\tau)$ is expressed in observer's units of Jy~Hz. 

The delay spectrum has a close resemblance to cosmological H~{\sc i} spatial power spectrum. Appendix~\ref{sec:EoR-power-spectrum} gives an overview of the similarities and differences between the two. Foregrounds can be described in either framework. For our study, we find the delay spectrum approach to be simple and yet extremely useful. 

In order to express a quantity derived from $\tilde{V}_b(\tau)$ whose units are the same as that of the cosmological H~{\sc i} power spectrum, we define the delay power spectrum:
\begin{align}\label{eqn:delay-power-spectrum}
  P_\textrm{d}(\boldsymbol{k}_\perp,k_\parallel) &\equiv |\tilde{V}_b(\tau)|^2\left(\frac{A_\textrm{e}}{\lambda^2\Delta B}\right)\left(\frac{D^2\Delta D}{\Delta B}\right)\left(\frac{\lambda^2}{2k_\textrm{B}}\right)^2,
\end{align}
with
\begin{align}
  \boldsymbol{k}_\perp &\equiv \frac{2\pi(\frac{\boldsymbol{b}}{\lambda})}{D}, \\
  k_\parallel &\equiv \frac{2\pi\tau\,f_{21}H_0\,E(z)}{c(1+z)^2}, 
\end{align}
where, $A_\textrm{e}$ is the effective area of the antenna, $\Delta B$ is the bandwidth, $\lambda$ is the wavelength of the band center, $k_\textrm{B}$ is the Boltzmann constant, $f_{21}$ is the rest frame frequency of the 21~cm spin flip transition of H~{\sc i}, $z$ is the redshift, $D\equiv D(z)$ is the transverse comoving distance, $\Delta D$ is the comoving depth along the line of sight corresponding to $\Delta B$, and $h$, $H_0$ and $E(z)\equiv [\Omega_\textrm{M}(1+z)^3+\Omega_\textrm{k}(1+z)^2+\Omega_\Lambda]^{1/2}$ are standard terms in cosmology. Throughout the paper, we use $\Omega_\textrm{M}=0.27$, $\Omega_\Lambda=0.73$, $\Omega_\textrm{K}=1-\Omega_\textrm{M}-\Omega_\Lambda$, $H_0=100\,$km$\,$s$^{-1}\,$Mpc$^{-1}$, and $P_\textrm{d}(\boldsymbol{k}_\perp,k_\parallel)$ is in units of K$^2$(Mpc/$h$)$^3$.

In summary, the delay spectrum, $\tilde{V}_b(\tau)$, is obtained from {\it visibilities} which are the basic data blocks measured by each antenna pair, using equations~\ref{eqn:obsvis} and \ref{eqn:delay-transform}. $\tilde{V}_b(\tau)$ captures all the effects of EoR H~{\sc i} signal corruption caused by foregrounds and the instrument. At the same time, it is closely related to the sought power spectrum containing critical information about spatial scales. In using the delay spectrum technique, visibilities from different baselines have not been averaged together. 

\subsection{Delay Space}\label{sec:delay-space}

\begin{figure}[htb]
\centering
\includegraphics[width=\linewidth]{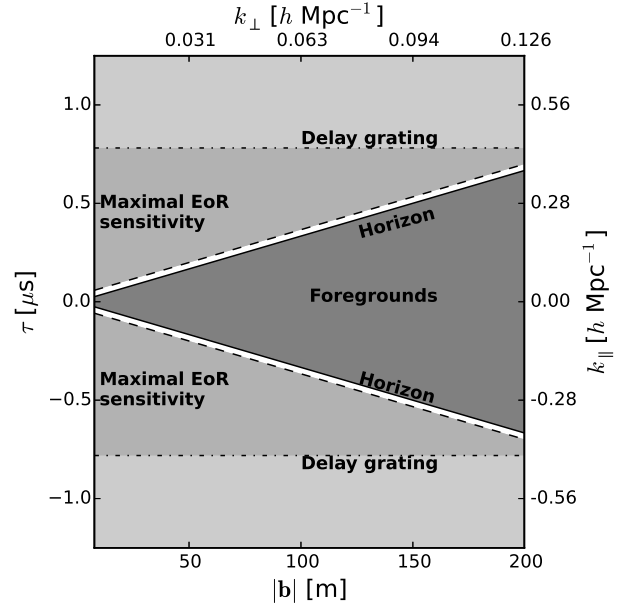}
\caption{Fourier space in which delay (and power) spectra of EoR H~{\sc i} signals are calculated. The $x$-axis is denoted by $|\boldsymbol{b}|$ (antenna spacing) or $k_\perp$ (transverse wavenumber). The $y$-axis denoted by $\tau$ (delay) or $k_\parallel$ (line of sight wavenumber). Here, $k_\perp$ and $k_\parallel$ are obtained for a frequency of 185~MHz. The dark shaded region is referred to as the {\it foreground wedge} where smooth spectrum foregrounds reside. Its boundaries (solid lines), given by light travel time for corresponding antenna spacings, are referred to as horizon delay limits. Narrow extensions of the {\it wedge} (white unshaded strips) are caused by convolution with the instrument's spectral transfer function. Regions excluding the {\it wedge} are expected to be relatively free of foreground contamination and are generally referred to as the {\it EoR window}. There are undesirable grating responses (dotted-dashed lines) specific to the MWA. Hence, we conservatively identify a restricted region of high EoR sensitivity (medium shade) and refer to it as the MWA {\it EoR window}. \label{fig:fourier-space}}
\end{figure}

We give a brief overview of some parameters of Fourier space which are generic to all experiments that use a similar approach. Figure~\ref{fig:fourier-space} illustrates the Fourier space in which the delay (and power) spectra of redshifted H~{\sc i} observations are calculated. $|\boldsymbol{b}|$ and $k_\perp$, denoting spatial scales in the transverse direction (tangent plane to the celestial sphere), form the $x$-axis. $\tau$ and $k_\parallel$, denoting spatial scales along line of sight form the $y$-axis. Foreground emission maps to a wedge-shaped region in Fourier space, hereafter referred to as the {\it foreground wedge} \citep{dat10}, whose boundaries are determined by the antenna spacings and the light travel times across them. These boundaries, called horizon delay limits \citep{ved12,par12b}, are shown by solid lines. 

The spectral transfer function of the instrument convolves the {\it foreground wedge} and stretches it further (unshaded narrow strips bounded by solid and dashed lines) along $\tau$-axis \citep{par12b,thy13}. The width of this narrow strip is inversely proportional to the operating bandwidth. The region of Fourier space excluding the {\it foreground wedge} and the narrow strips is the so-called {\it EoR window}, shown in light and medium shades. In the context of EoR studies in Fourier space, the H~{\sc i} power spectrum from the EoR is expected to decrease rapidly with $|\boldsymbol{k}|$. Hence, the brightest EoR signal will be observed on the shortest baselines and smallest delays. Thus the regions of interest for EoR studies relying on {\it avoidance} strategy are just beyond the horizon delay limits (dashed lines) on short baselines, marked as regions of {\it maximal EoR sensitivity}. 

In the specific case of the MWA, which has a passband constructed using coarse channels, there are period grating responses resulting in repetitions of the {\it foreground wedge} at multiples of 0.78~$\mu$s. Thus the MWA {\it EoR window} lies outside the dashed lines but inside the first grating response ($|\tau| \lesssim 0.78\,\mu$s, dotted-dashed lines) and is shown in medium shade.

\subsection{Delay Spectrum Deconvolution}\label{sec:data-delay-spectrum}

We obtain the delay spectrum of visibilities by taking the delay transform of each baseline's spectrum (equation~\ref{eqn:delay-transform}) choosing $W(f)$ to be a {\it Blackman-Harris} window function. The sky spectrum is multiplied in the instrument by the instrumental passband and flagging of frequency channels possibly corrupted by radio frequency interference, which together are represented by the weights, $W_\textrm{i}(f)$. In delay-space, these weights translate into a convolution by a point spread function (PSF). We deconvolve this PSF using a one dimensional CLEAN algorithm \citep{tay99} as described for the delay axis \citep{par09,par12b} to obtain the final delay spectra. The CLEAN procedure iteratively finds and subtracts peak values convolved by the Fourier transform of the weights. We limit the selection of peaks to modes inside the horizon delay limit, corresponding to smooth spectrum objects in the visible sky hemisphere. 

\section{Wide-Field Measurements}\label{sec:wide-field}

With $\hat{\boldsymbol{s}}\equiv (l,m,n)$, equation~\ref{eqn:obsvis} can be written as \citep{tay99,tho01}:
\begin{align}\label{eqn:obsvis-lmn}
  V_b(f) &= \iint\limits_\textrm{sky} \frac{A(\hat{\boldsymbol{s}},f)\,I(\hat{\boldsymbol{s}},f)}{\sqrt{1-l^2-m^2}}\,W_\textrm{i}(f)\,e^{-i2\pi f\frac{\boldsymbol{b}\cdot\hat{\boldsymbol{s}}}{c}}\,\dif l\,\dif m, 
\end{align}
where, $l$, $m$, and $n$ denote the direction cosines toward east, north, and zenith respectively, with $n\equiv\sqrt{1-l^2-m^2}$, and:
\begin{align}\label{eqn:solid-angle}
  \dif\Omega = \frac{\dif l\,\dif m}{\sqrt{1-l^2-m^2}}.
\end{align}

When the synthesized field is small, where $A(\hat{\boldsymbol{s}},f)$ or $I(\hat{\boldsymbol{s}},f)$ is significant only for $|l| \ll 1$ and $|m| \ll 1$, equation~\ref{eqn:obsvis-lmn} reduces to a simple two-dimensional Fourier transform \citep{tay99,tho01} between the apparent sky brightness and measured visibilities. It is in this context that radio interferometers are understood to be sensitive only to fluctuations and not to a uniform sky brightness distribution. 

In a wide-field measurement, neither $A(\hat{\boldsymbol{s}},f)$ nor $I(\hat{\boldsymbol{s}},f)$, in general, is negligible anywhere in the visible hemisphere. The solid angle per pixel on the sky in direction cosine coordinates changes significantly with direction (equation~\ref{eqn:solid-angle}), increasing steeply toward the horizon. Hence, the approximations in the narrow-field scenario do not apply. For example, even if $A(\hat{\boldsymbol{s}},f)$ and $I(\hat{\boldsymbol{s}},f)$ are held constant across the visible hemisphere, the amplitude of the integrand in equation~\ref{eqn:obsvis-lmn} is still dependent on direction. Therefore, in a significant departure from a narrow-field measurement, the wide-field visibility from a uniform brightness distribution on a non-zero antenna spacing is not zero.

\begin{figure}[htb]
\centering
\includegraphics[width=\linewidth]{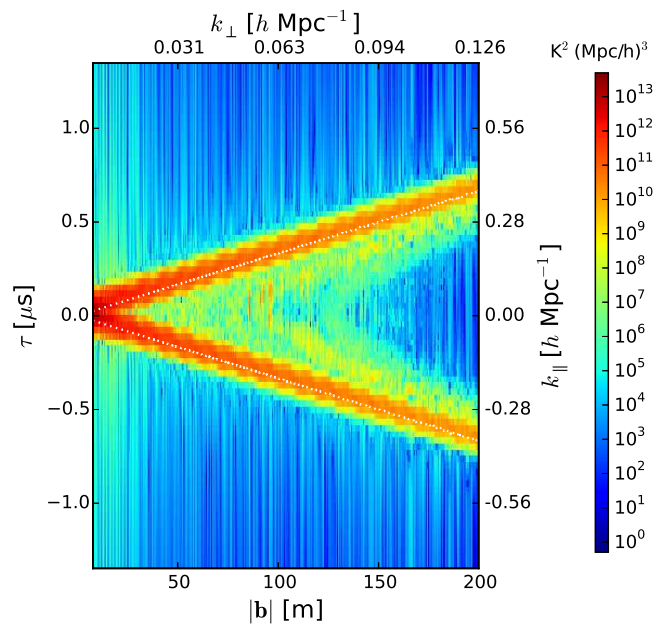}
\caption{Wide-field effects on delay power spectra produced with a uniform sky brightness distribution measured by antenna pairs with a uniform power pattern across the visible hemisphere. Delay power spectra are obtained using equations~\ref{eqn:obsvis}, \ref{eqn:delay-transform} and \ref{eqn:delay-power-spectrum} for each baseline, which are then stacked by baseline length. The axes correspond to cosmological dimensions. The non-zero response of the interferometer array to a uniform brightness distribution and the prominent {\it edge brightening} close to the horizon delay limits are wide-field effects. These are prevalent on all antenna spacings and are generic to all instruments used in wide-field measurements. \label{fig:usm}}
\end{figure}

Figure~\ref{fig:usm} shows the wide-field delay power spectrum of a uniformly illuminated sky with no spectral variation as measured by antenna elements arranged identical to that in the MWA antenna array layout \citep{bea12} with a uniform power pattern across the sky and a bandwidth of 30.72~MHz centered around 185~MHz (refer to \S\ref{sec:instrument_model} for a detailed description of the instrument model). Notice the steep rise in power toward the horizon limits. These wide-field effects are prevalent on all antenna spacings, including the longest ones used in this study. 

We interpret this as due to equal-sized delay bins subtending larger solid angles near the horizon thereby containing larger integrated emission. Further, baseline vectors (including those with largest lengths) are foreshortened toward the horizon along their orientation. Thus, they become sensitive to larger angular scales that match the inverse of their foreshortened lengths along these directions.

\citet{thy13} found evidence of this feature in their statistical models. In line with our reasoning, they attribute it to a steep rise in solid angles subtended by delay bins near the horizon limits. \citet{pob13} also find a similar ``edge brightening'' feature which they attribute to Galactic plane emission near the horizon. From their discussion, it is unclear what fraction of power in that feature arises from such wide-field effects.

We conclude these are generic to all instruments making wide-field measurements. The nature of the specific instrument used for observing will control the amplitude of these effects, which we explore below. 

\section{Simulations}\label{sec:sim}

We describe the instrument and foreground models used in our simulations. 

\subsection{Instrument Model}\label{sec:instrument_model}

In our present study, we use a latitude of -26\fdg 701 and an antenna layout identical to that of the MWA \citep{bea12} for the observatory. The array is arranged as a centrally condensed core of $\sim$~300~m --- there are many spacings in the range 5--50~m --- and a radial density that falls off as the inverse of the radius, with the longest baselines at 3~km. Here we focus on antenna spacings $|\boldsymbol{b}| \le 200$~m (spatial scales relevant to reionization). Their deviation from coplanarity is negligible. For geometrical intuition, we restrict the orientation ($\theta_b$, measured anti-clockwise from east) of all baselines to lie in the range $-67\fdg 5 \leq \theta_b < 112\fdg 5$. Baselines oriented in the other half-plane measure conjugate visibilities with delays of equal magnitude but of opposite sign and hence are ignored in our analysis. We choose an observing frequency of 185~MHz ($z\simeq 6.68$) and a flat passband of width $\Delta B = 30.72$~MHz to roughly match those of ongoing MWA EoR observations (discussed in detail in \S\ref{sec:MWA}). 

One of the principal components of the instrument model is the antenna power pattern, $A(\hat{\boldsymbol{s}},f)$ (see equation~\ref{eqn:obsvis}). It is determined by the shape of its aperture. Using a few examples, we examine the role the geometrical shape of the aperture plays in shaping the characteristics of delay power spectrum. We consider the following antenna elements placed at the MWA tile locations:
\begin{enumerate}
\item {\it Dipole}: an east-west dipole of length 0.74~m at a height 0.3~m above a ground plane. $A_\textrm{e}=(\lambda/2)^2$.
\item {\it Phased Array}: a 4$\times$4 array of isotropic radiators with a grid spacing of 1.1~m at a height 0.3~m above the ground plane placed in an arrangement similar to that of an MWA tile. $A_\textrm{e}=16\,(\lambda/2)^2$. 
\item {\it Dish}: diameter of 14~m similar to that proposed for HERA, with $A_\textrm{e}\approx 154$~m$^2$. The power pattern is simulated using an {\it Airy} pattern where its sensitivity beyond the horizon is forced to zero.
\end{enumerate}

\subsection{Foreground Model}\label{sec:foreground}

In wide-field measurements, it is important to consider an all-sky model for foreground objects in evaluating the features seen in the power spectrum instead of restricting only to the primary field of view, a point also supported by Pober et al. (2015, in preparation). We use a foreground model that includes both diffuse and bright compact components. 

For the diffuse component, we use an all-sky radio foreground model \citep{deo08} to estimate the emission at 185~MHz. At this frequency, since this map is predominantly based on the 408~MHz map of \citet{has82} which has an angular resolution of 0\fdg 85, we smoothed the 185~MHz map to the same resolution. However, to avoid any artifacts from sampling this map, we sample it at $\approx 27$\arcmin~intervals. We model the diffuse foreground spectra with a unique spectral index at each pixel in the map, estimated from model maps at 170~MHz and 200~MHz.

The model described above is primarily a model of the diffuse foreground sky. While it contains faint compact emission blended in with the diffuse emission, bright point sources have been removed \citep{deo08}. In order to supplement it with missing bright compact emission, we use classical radio source confusion estimates to determine the nominal flux density threshold and include point sources brighter than this threshold. Slightly different criteria are in common use in radio astronomy to estimate radio source confusion \citep[see Appendix of][and references therein]{thy13}. For an angular resolution of 0\fdg 85, using a conservative `$S_\textrm{c}=5\sigma_\textrm{c}$' criterion, we determine the flux density threshold to be $\approx 10$~Jy. Other liberal criteria that yield a lower threshold carry a greater risk of double-counting point sources which might be already blended in with the diffuse sky model. 

We use a combination of the NRAO VLA Sky Survey \citep[NVSS;][]{con98} at 1.4~GHz and the Sydney University Molonglo Sky Survey \citep[SUMSS;][]{boc99,mau03} at 843~MHz to provide our point source catalog due to their complementary survey footprints covering the entire sky, and matched flux density sensitivity and angular resolution. The SUMSS catalog covers the sky with declination $\delta < -30\arcdeg$~ with a limiting peak brightness of 6--10~mJy/beam and an angular resolution of $\sim 45$\arcsec. The NVSS covers the sky with $\delta > -40\arcdeg$~ with a similar angular resolution and a limiting flux density of $\approx 2.5$~mJy for point sources. 

From the SUMSS catalog, we select objects whose deconvolved major axes are equal to 0\arcsec, thereby strictly selecting point sources. From the NVSS catalog, we excluded objects that overlap with those in the SUMSS survey footprint. Point sources from NVSS were selected if the convolved major axes were not greater than $\approx 47$\arcsec, which matches the angular resolution of the survey. Using a mean spectral index of $\langle\alpha_\textrm{sp}\rangle=-0.83$ (flux density, $S(f)\propto f^{\alpha_\textrm{sp}}$) obtained by \citet{mau03} for both NVSS and SUMSS catalog objects, we calculate the corresponding flux densities at 185~MHz, $S_{185}$. From this subset, we choose point sources with $S_{185}\geq 10$~Jy. The selection of such bright point sources is not affected by minor differences in flux density sensitivity of the two surveys. We verified that our selection criteria ensure a similar areal density of objects in the two surveys. 

\begin{figure}[htb]
\centering
\includegraphics[width=\linewidth]{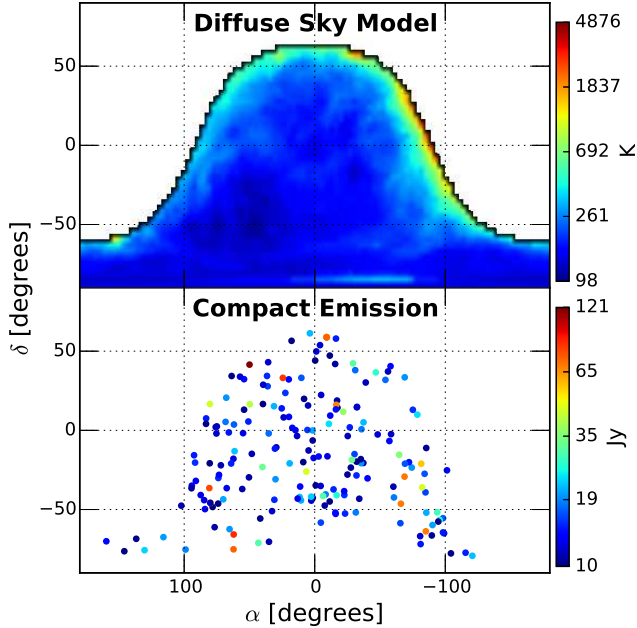}
\caption{Foreground model at 185~MHz consisting of diffuse emission (top) in units of K and bright point sources (bottom) in units of Jy, visible during a snapshot at 0.09~hr LST. In the diffuse model, the Galactic center has just set in the west. Color scales are logarithmic. \label{fig:sky-model-generic}}
\end{figure}

These criteria yield 100 objects from the SUMSS catalog and 250 objects from the NVSS catalog. Together with the diffuse foreground model, we obtain an all-sky foreground model consisting of both compact and diffuse emission. Figure~\ref{fig:sky-model-generic} shows the diffuse (top) and compact (bottom) foreground emission model used in our study. In this snapshot pointed toward zenith at 0.09~hr LST, the Galactic center in the diffuse model has just set in the west. 

\subsection{Role of Antenna Geometry}\label{sec:antenna-shape}

\begin{figure*}[htb]
\centering
\subfloat[][Power patterns]{\label{fig:power-patterns}\includegraphics[width=\linewidth]{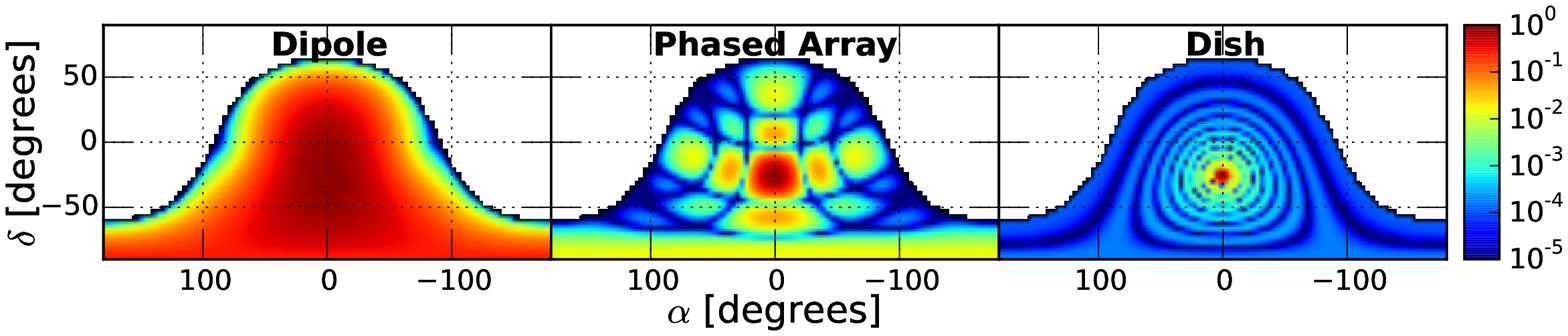}} \\
\subfloat[][Delay power spectra]{\label{fig:antennas-delay-spectra}\includegraphics[width=\linewidth]{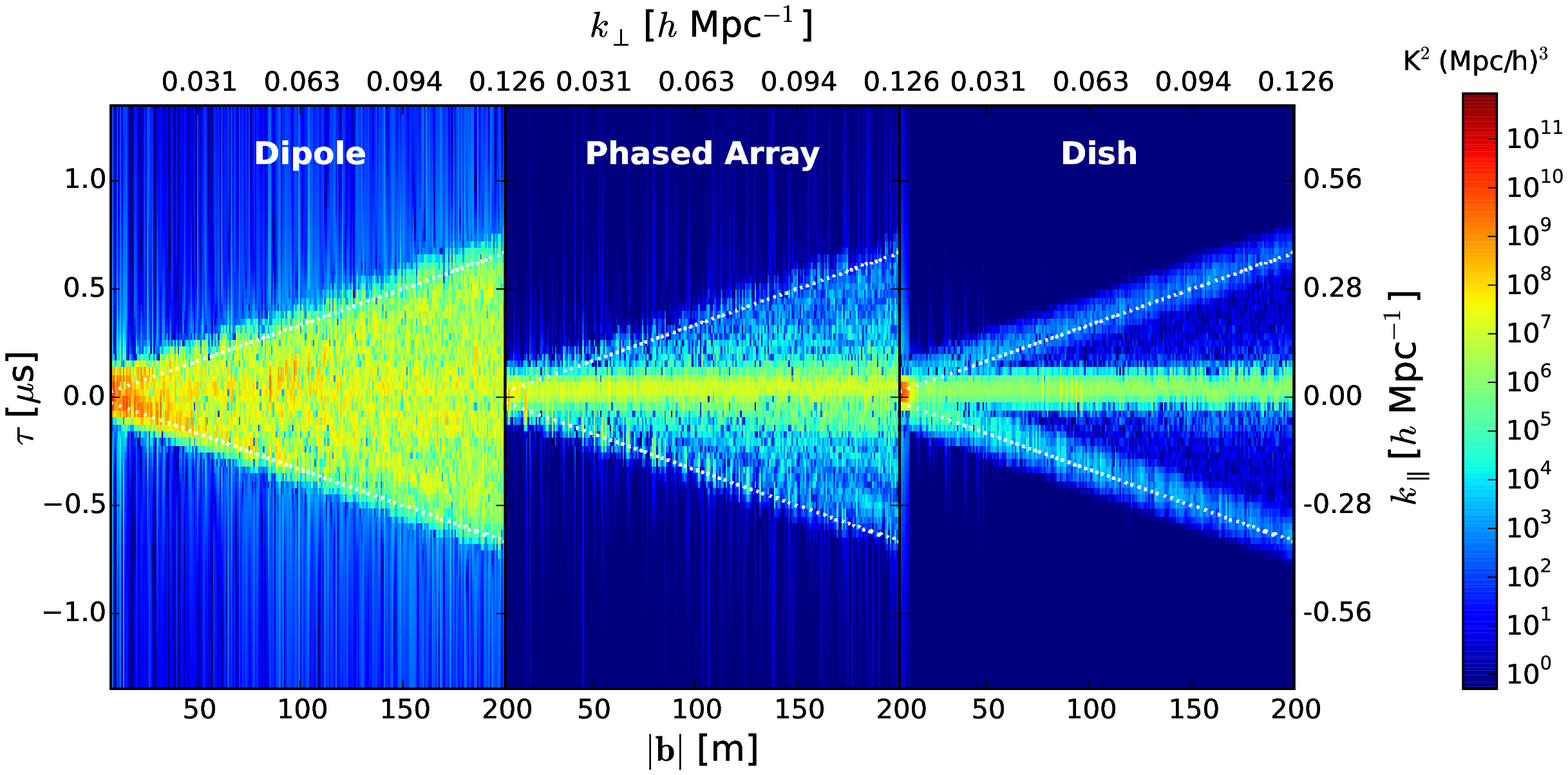}}
\caption{Power patterns (top panels) and simulated delay power spectra (bottom panels) for different antenna shapes at 185~MHz centered on zenith. Antenna shapes used are: dipole (left), phased array (middle), and dish (right). Refer to \S\ref{sec:instrument_model} for details of the antenna models. The strength and occupancy of the power patterns are correlated with those of delay power spectra. White dotted lines in the delay power spectra mark the boundaries of the {\it foreground wedge} determined by the horizon delay limit and antenna spacing. The {\it foreground wedge} and the {\it EoR window} are most severely contaminated in the case of the dipole while it is the least for the dish. The phased array has intermediate levels of contamination. Foreground power close to the horizon delay limits in all three cases is significant even on long baselines. The foreshortening of baselines toward the horizon makes them sensitive to foreground emission on large size scales. The amplitude of this feature strongly depends on the shape of the antenna element. It is highest for a dipole (which has a strong response near the horizon) and least for a dish.}
\label{fig:aperture-shapes}
\end{figure*}

The power patterns of the aforementioned antenna geometries at 185~MHz for this zenith pointing are shown in Figure~\ref{fig:power-patterns}. 

The delay power spectra without thermal noise component for these antenna shapes are shown in Figure~\ref{fig:antennas-delay-spectra}. The occupancy of the power patterns on the sky is clearly correlated with that in the delay spectra. Further, the strength of the primary lobe centered on the pointing center is correlated with the delay power spectrum centered on $\tau=0$; and, the overall rate of decrease in the power sensitivity away from the pointing center is correlated with the rate of drop in power away from $\tau=0$. 

The levels of foreground contamination in the {\it EoR window} varies substantially across the different antenna shapes: $\sim 10^4\,$K$^2\,$(Mpc/$h$)$^3$, $\lesssim 10^2\,$K$^2\,$(Mpc/$h$)$^3$, and $<1\,$K$^2\,$(Mpc/$h$)$^3$ for the dipole, phased array, and dish, respectively. The severity of foreground contamination inside the {\it foreground wedge} both in strength and occupancy also evidently decreases as the antenna element is changed from a dipole to a phased array to a dish. For instance, notice that the foreground contamination in $k$-modes between $k_\parallel=0$ and the horizon limits decreases from $\sim 10^5$~K$^2\,$(Mpc/$h$)$^3$ in a phased array to $\sim 10$~K$^2\,$(Mpc/$h$)$^3$ in a dish. As a consequence, $k$-modes in the {\it foreground wedge} that may be deemed too contaminated for EoR studies in the case of a dipole or a phased array can potentially become accessible when using a dish.

Finally, a distinct feature common to all these aperture shapes is that the foreground contamination near the horizon delay edges is significant even on wide antenna spacings ($\gtrsim 10^5$~K$^2$(Mpc/$h$)$^3$). We have argued this arises due to wide-field effects. The prevalence of this feature across different antenna shapes demonstrates it is generic to all wide-field measurements. The amplitude of this effect, however, can be controlled via choice of antenna shape and through weighting of aperture illumination. A dish-shaped antenna appears to hold a significant advantage over a dipole or a phased array from the viewpoint of foreground contamination. 

Typically, the sensitivity of antennas to the primary field of view is high compared to the rest of regions on the sky. Combined with the wide-field effects seen earlier, it leads to a ``pitchfork''-shaped signature inside the {\it foreground wedge}, as exemplified in the case of a dish. Although the exact appearance of this signature depends on the antenna power pattern, we use the term {\it pitchfork} hereafter, to broadly refer to the combination of foreground power in the primary field of view and the enhancement of foreground power near the horizon limits due to the nature of wide-field measurements.

\section{The Murchison Widefield Array}\label{sec:MWA}

We now use our simulations to analyze features in observed delay power spectrum obtained using the MWA instrument \citep{lon09,tin13}.

MWA construction was completed in 2012 and, after commissioning, began its EoR observing program in 2013. The MWA is a 128-tile interferometer capable of observing a 30.72~MHz instantaneous band anywhere in the range 80--300~MHz. Each tile is a phased array of 16 dipoles, each in the shape of a bow-tie. This yields a primary field of view $\gtrsim$~20\arcdeg~ wide and multiple secondary lobes. See \citet{bea12} for the tile layout. 

The MWA passband of width $\Delta B=30.72$~MHz is divided coarsely into 24$\times$1.28~MHz sub-bands with each sub-band weighted by a digital filter. The coarse channel shape is obtained using an eight-tap polyphase filter bank (PFB) and a Kaiser window with parameter $\beta=5$. Each of these coarse bands consists of 32 fine channels of width 40~kHz each. After correcting for the shape of these coarse channels, the fine channels at the edges of these sub-bands are flagged because they are known to be contaminated by aliasing at a low level. 

The MWA is expected to be sensitive to the power spectrum of the H~{\sc i} signal over the redshift range $6<z<10$ \citep{bow06,bea13,thy13}. Over 600~hr have been currently observed using the MWA, targeting science objectives outlined in \citet{bow13}.

The MWA targets two primary low-foreground fields for reionization observations. Here, we focus on the field at R.A.~$=0^\textrm{h}$, decl.~$=-30\arcdeg$. The MWA tracks a patch of sky through antenna beams formed and steered electronically by controlling delay settings of the dipoles in an MWA tile. The pointing system is optimized to points on a regular $\sim$7\arcdeg~ grid. The sky drifts across to the nearest available pointing, shifting between grid points (once every $\sim 30$~minutes). This process is repeated throughout the course of the observation $\approx 4.86$~hr. 

The observations used here were acquired on 2013~August~23. We have chosen two sections of duration 112~s each from this night for detailed study. These were chosen to provide a selection of possible foreground and instrumental conditions. As an example of a nominal observing setup we choose a zenith pointing; as an example of poor foreground conditions, we choose a pointing when the field is $\sim 2$~hr from zenith. This pointing has a significantly higher secondary lobe structure and is observed when the bright galactic center is well above the horizon. 

These two pointings are at LST 22.08~hr and 0.09~hr, which are hereafter denoted as {\it off-zenith} and {\it zenith} pointings, respectively. 

\subsection{Initial Data Processing}\label{sec:data-analysis}

The data are flagged for interference \citep{off15}, removing 3\% of the data and averaged in time  and frequency from the raw 0.5~s, 40~kHz to 2s, 80~kHz. These data are then calibrated to a simulation of the sky containing 2420 point-like objects selected from the MWA Commissioning Survey \citep[MWACS;][]{hur14}. It has a flux density limit of 25~mJy and a declination range of $-12\arcdeg$~ to $-40\arcdeg$~ evenly covering the field of view of the observations reported here. The objects used in calibration are selected to lie inside the 5\% contour of the primary lobe of the tile power pattern. The calibration algorithm --- based on forward modeling software by \citet{sul12} and the calibration method described by \citet{sal14} --- computes complex gain solutions per channel per antenna averaged to two minute intervals. The solutions are fairly low signal to noise so we iteratively average along the antenna and frequency dimensions to capture the relatively independent passband and antenna--to--antenna variation. First, we average the channel gains over all antennas to obtain a high signal to noise measurement of the bandpass. After applying this single passband, we do a second round of calibration and fit second and first order polynomials for amplitude and phase respectively for each antenna. This flattens any residual variation in bandpass and removes small phase slopes due to variations in cable delay. Finally, we fit for an additional phase known to be caused by small reflections in a subset of cables. 

\subsection{Modeling}\label{sec:modeling}

The MWA tile power pattern is modeled as a mutually-coupled 4-by-4 dipole array with the overall power pattern of each individual dipole calculated via finite element electromagnetic simulation \citep{sut14_arXiv}. To speed up simulations, we find that a phased array of isotropic radiators at a height of 0.3~m above an infinite ground plane provides a very good approximation to the full simulation, hence we use the idealized dipoles. We also assume that each individual dipole signal has random delay fluctuations of rms 0.05~ns, a number in line with the known repeatability and stability level of the analog signal chain \citep{bow07b}. Besides having the effect of adding a time-dependent uncertainty in the power pattern, these random delay fluctuations reduce the coherence in the phased addition of dipole signals resulting in deviations from predicted models of the power pattern, most prominently at its nulls. 

\begin{figure*}[htb]
\centering
\subfloat[][Diffuse Sky Model]{\label{fig:DSM}\includegraphics[width=0.45\linewidth]{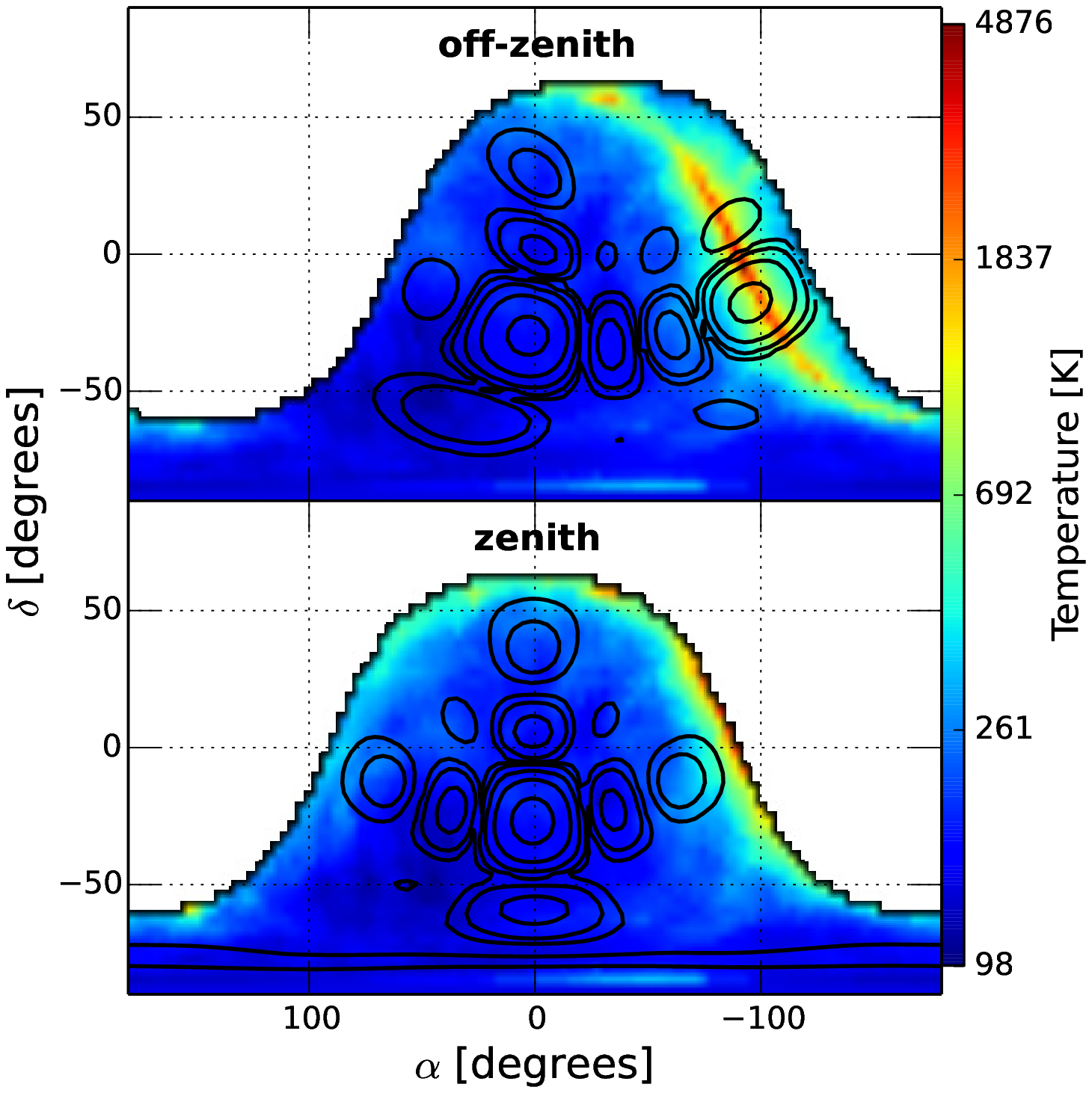}}
\subfloat[][Bright Point Source Model]{\label{fig:CSM}\includegraphics[width=0.45\linewidth]{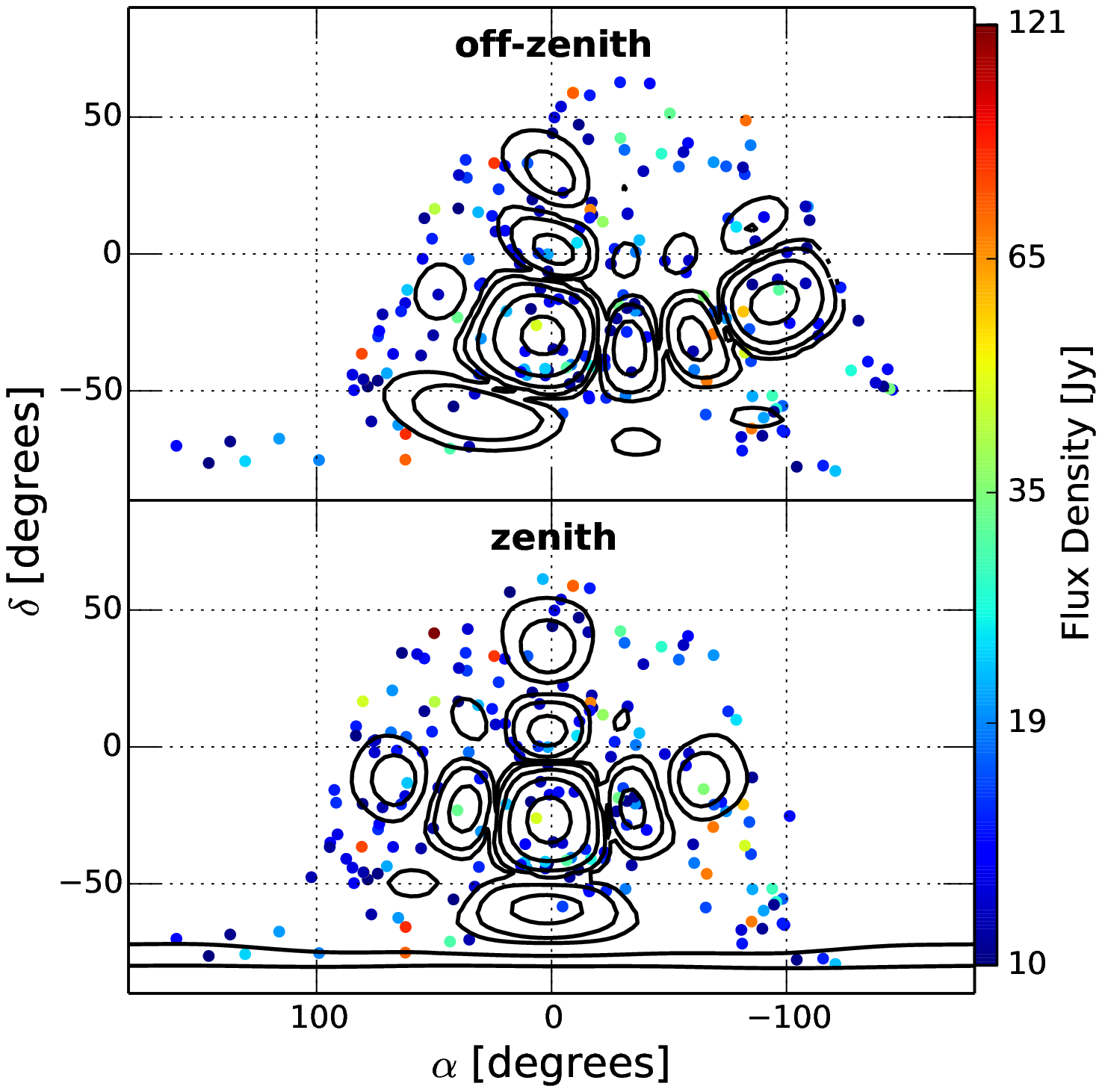}}
\caption{Sky brightness temperature of the diffuse foreground model (left) and flux densities of bright point sources (right) at 185~MHz visible during {\it off-zenith} (top) and {\it zenith} (bottom) pointings. The color scales are logarithmic. MWA tile power pattern contours are overlaid. The contour levels shown are 0.00195, 0.00781, 0.0312, 0.125, and 0.5. The Galactic center and a portion of the Galactic plane are prominently visible during the {\it off-zenith} pointing in the diffuse sky model and the MWA tile power gain is significant ($\gtrsim 12$\%) in that direction. In contrast, emission from the Galactic plane in {\it zenith} pointing is significantly less.}
\label{fig:sky-model-MWA}
\end{figure*}

We use the model described in \S\ref{sec:foreground} for the foreground sky. Figure~\ref{fig:sky-model-MWA} shows the diffuse emission and bright point source foreground models for the two chosen pointings with the modeled MWA tile power pattern contours overlaid. Notice the presence of a portion of the Galactic plane and the bright Galactic center in the westward sky in the diffuse sky model, where the MWA tile power gain is significant ($\gtrsim 12$\%). In the {\it zenith} pointing, the Galactic plane has set and the power pattern in that direction is at least 16 times smaller. 

We estimate thermal noise, $\Delta V$, in the observed visibilities, $V_b(f)$, using the rms of $\tilde{V}_b(\tau)$ obtained from data after {\it delay-deconvolution} across all antenna spacings for $|\tau| \geq 1$~$\mu$s using the relations:
\begin{align}\label{eqn:Tsys}
  \Delta \tilde{V} &= \sqrt{N_\textrm{ch}}\,\Delta V \Delta f, \quad\textrm{and} \\
  \Delta V &= \frac{2\,k_\textrm{B}\,T_\textrm{sys}}{A_\textrm{e}\sqrt{2\,\Delta f\,\Delta t}},
\end{align}
where, $\Delta f = 80$~kHz, $\Delta t = 112$~s, and $N_\textrm{ch}=\Delta B/\Delta f$ is the number of frequency channels. The choice of threshold for $\tau$ is well outside the {\it foreground window}, where foreground contamination is negligible and thus yields a robust estimate of $T_\textrm{sys}$. We find the average system temperature to be $\sim 95$~K. Hence, for our simulations, we use $T_\textrm{sys}=95$~K to match the thermal noise in data. 

\subsection{Comparison of Data and Model}\label{sec:data-vs-model}

\begin{figure*}[htb]
\centering
\includegraphics[width=\linewidth]{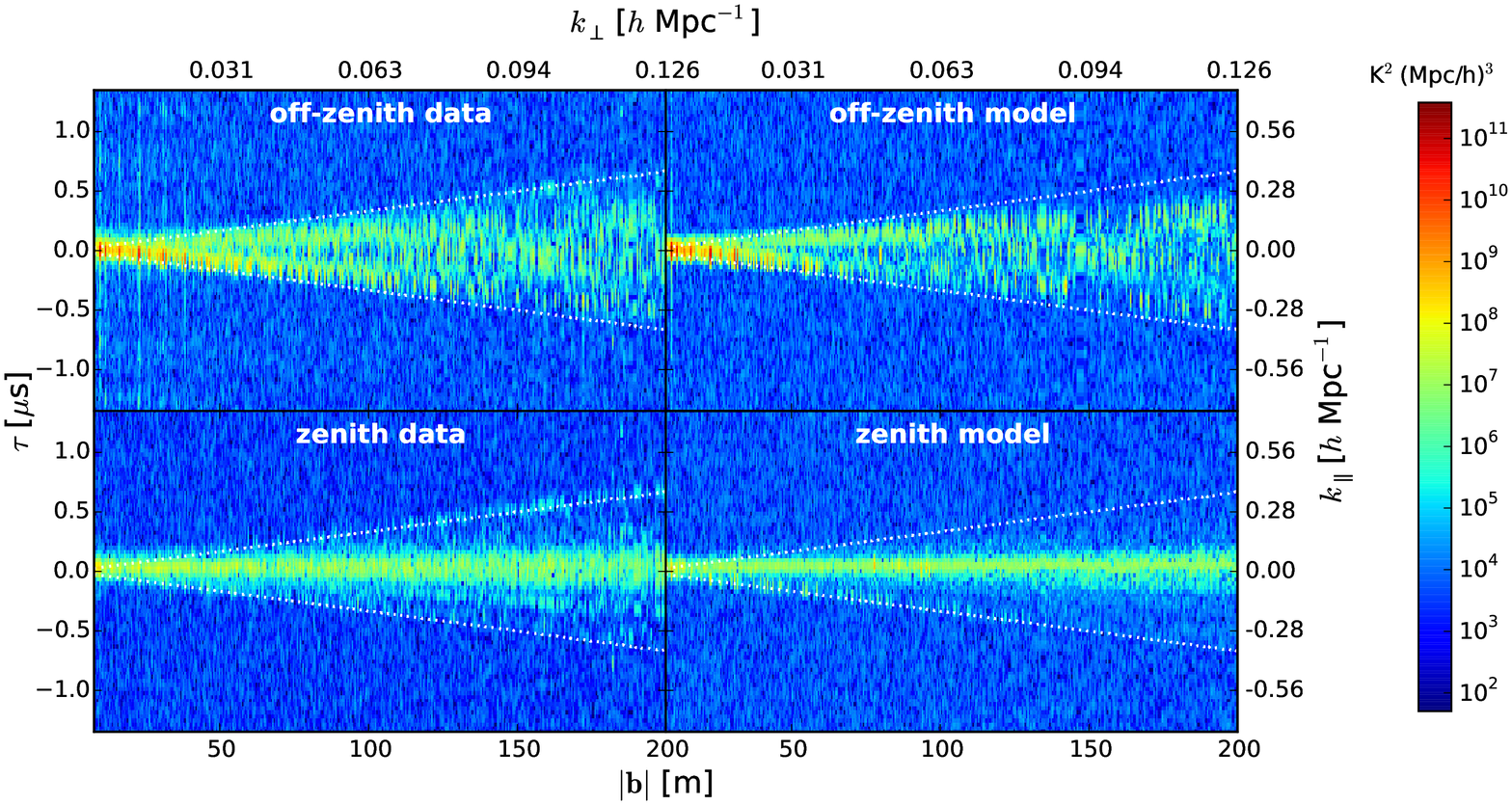}
\caption{Delay power spectra from MWA data (left) and modeling (right) for the {\it off-zenith} (top) and {\it zenith} (bottom) pointings. The {\it foreground wedge} is bounded by white dotted lines. Model matches the data to a level consistent with the uncertainties in foreground models and the antenna beam (as discussed in \S\ref{sec:data-vs-model}). \label{fig:fhd-sim-comparison}}
\end{figure*}

With the aforementioned foreground model, and instrumental and observational parameters, we simulate visibilities using equation~\ref{eqn:obsvis}. Figure~\ref{fig:fhd-sim-comparison} shows the delay power spectra from {\it off-zenith} and {\it zenith} pointings obtained from MWA observations and modeling. Notice the qualitative agreement of amplitude and structure between the two. The Galactic center and the Galactic plane visible in the {\it off-zenith} pointing make it appear brighter in the {\it foreground wedge} as a branch with $\tau<0$. 

In order to make a quantitative comparison of delay spectra obtained with MWA data and our simulations, we consider the uncertainty in the assumed spectral index of our foreground model. Our foreground models are derived from other higher frequency catalogs and sky maps. The inherent spread in spectral index increases the uncertainty while predicting fluxes at the observing frequency. Using simple error propagation, the fractional error in the delay spectrum amplitude caused by the spread in spectral index is $\sim \ln(f_\textrm{orig}/f)\,\Delta\alpha_\textrm{sp}$, where, $f_\textrm{orig}$ is the original frequency at which the catalog or map was created, $f=185$~MHz is the MWA observing frequency, and $\Delta\alpha_\textrm{sp}$ is the spread (HWHM) in spectral index. From \citet{mau03}, we assume $\Delta\alpha_\textrm{sp} \approx 0.35$ for point sources from NVSS and SUMSS catalogs. Although the model of \citet{deo08} yields a spectral index per direction on the sky, we could assume similar uncertainties exist in spectral indices of our diffuse sky model as well, which is predominantly derived from the 408~MHz map of \citet{has82}. Thus, fractional errors in delay spectrum amplitudes from compact and diffuse components are $\sim$70\% and $\sim$30\% respectively. 

In addition to intrinsic model uncertainty, delay spectra from simulations and data each have fluctuations due to thermal noise in the delay spectrum with rms $\sim$ 1.4$\times 10^7$~Jy~Hz. We estimate the ratio of delay spectra from data and simulations as $\rho = |\tilde{V}^\textrm{D}_b(\tau)|\,/\,|\tilde{V}^\textrm{S}_b(\tau)|$, where superscripts D and S denote data and simulation, respectively. The median absolute deviation of $\log_{10}\rho$ inside the {\it foreground wedge} for both pointings is $\approx 0.28$. This corresponds to $\sim$~90\% fractional difference between data and modeling on average with either pointing. 

We also simulated delay spectra after assigning spectral indices drawn randomly from a {\it gaussian} distribution with a mean of $\langle\alpha_\textrm{sp}\rangle=-0.83$ and a HWHM of $\Delta\alpha_\textrm{sp}=0.35$ to the point sources in our compact foreground model. These simulations typically yielded a median absolute deviation of $\approx 0.29$ for $\log_{10}\rho$ indicating fractional differences of $\sim$~95\% between different realizations. This demonstrates that a fractional deviation of $\sim$~90\% observed between data and simulations is in line with expectations when the aforementioned uncertainty in foreground models, thermal noise fluctuations in measurements, and uncertainties in antenna power pattern due to random delay fluctuations are taken into account. 

These uncertainties are presented only to confirm the qualitative agreement already seen between data and modeling in Figure~\ref{fig:fhd-sim-comparison}. These estimates are conservative. A full treatment of all uncertainties and deviations from ideal behavior such as frequency dependent errors in tile power pattern \citep{ber15}, calibration \citep{dat10}, data corruption due to interference, anisoplanatic wide-field imaging and ionospheric effects \citep{int09} will bring the simulations much closer in agreement with observations, but is beyond the scope of this paper. Hereafter, our focus is to explore in detail the foreground signatures embedded in the {\it foreground wedge} of the MWA instrument.

\subsection{Analysis of Foreground Signatures}\label{sec:FG-analysis}

Having shown that the simulation matches the data to the level of expected uncertainties, we proceed to examine in further detail the key signatures seen in simulated delay spectra. A number of factors are responsible for the characteristics noted in the delay spectra obtained from data and through simulations. In subsequent sections, we provide a detailed explanation of our results as a combination of these factors. Note that numerous features may overlap at different degrees of significance depending on combinations of parameters. We assign the features to their predominant causes. Secondly, we have used noiseless cases to clearly illustrate the observed foreground signatures. With the addition of noise in the visibilities, some of the weaker features may not be as prominently visible. Since the foreground signatures are far too numerous and subject to a multitude of parameters like baseline length and orientation, power pattern, patch of sky under observation, and instrumental configuration, we highlight only the most notable features in the foreground delay power spectra.

\begin{figure*}[htb]
\centering
\includegraphics[width=\linewidth]{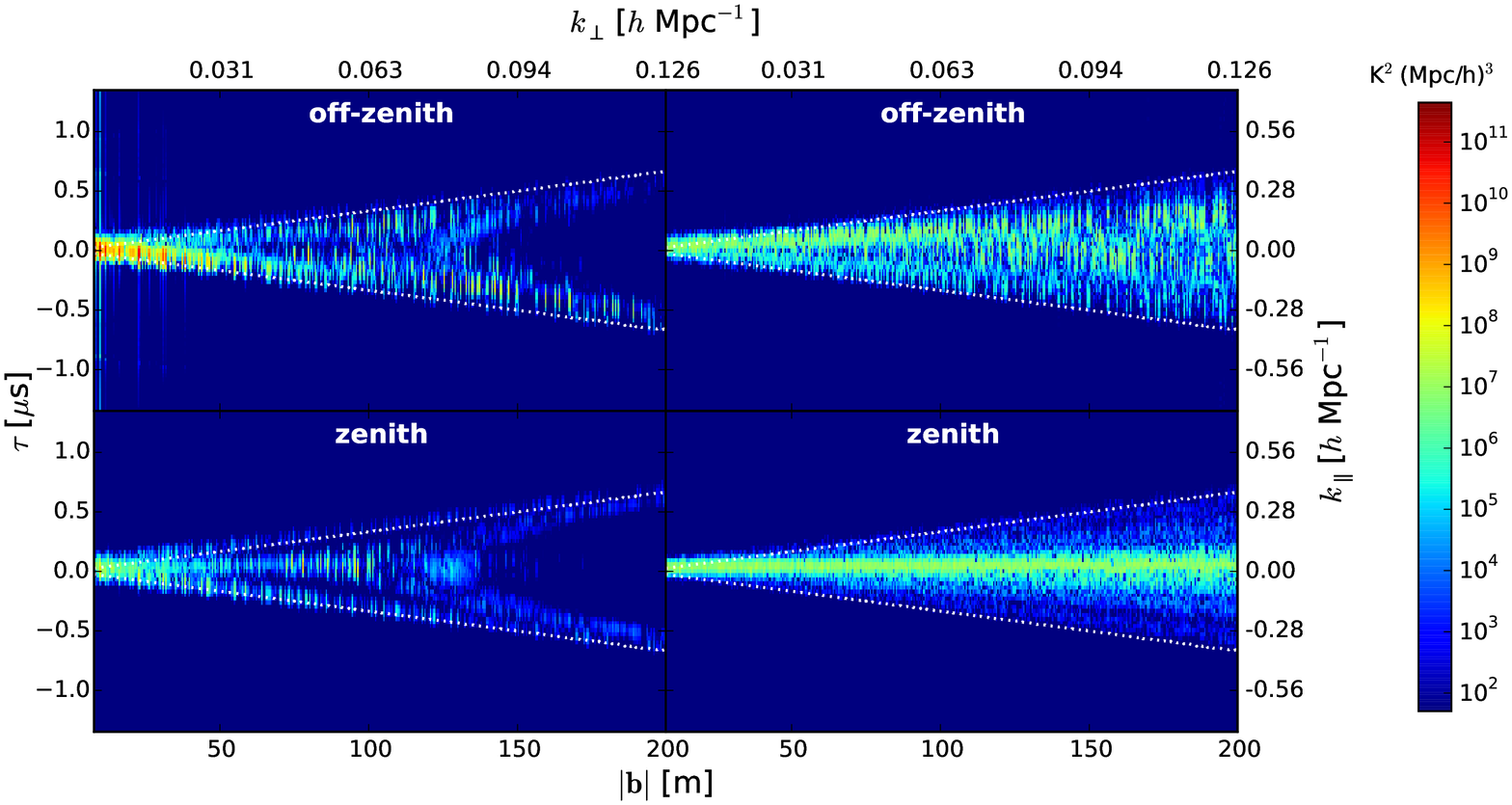}
\caption{Simulated delay power spectra (in units of K$^2$~(Mpc/$h$)$^3$) for for the diffuse (left) and compact (right) foreground models in the {\it off-zenith} (top) and {\it zenith} (bottom) pointings without any thermal noise. The axes and color scale are identical to those in Figure~\ref{fig:fhd-sim-comparison}. In the {\it off-zenith} pointing, emission from the Galactic center is the most prominent feature seen as a branch at $\tau<0$. In the {\it zenith} pointing, delay power spectrum from diffuse emission has a {\it two-pronged fork}-shaped structure and is present even at wide antenna spacings due to wide-field effects. Compact emission is centrally concentrated. \label{fig:noiseless-dsm-csm-delay-spectrum}}
\end{figure*}

Figure~\ref{fig:noiseless-dsm-csm-delay-spectrum} shows the delay power spectra obtained from the diffuse (left) and compact (right) foreground emission for the {\it off-zenith} (top) and {\it zenith} (bottom) pointings without thermal noise component. Some of the notable signatures are discussed below.

\subsubsection{Galactic Center on Eastward Antenna Spacings}\label{sec:GC-east}

The most prominent signature seen in the {\it off-zenith} pointing (top left panel, Figure~\ref{fig:noiseless-dsm-csm-delay-spectrum}) is due to the bright Galactic center situated on the western horizon co-located with one of the bright secondary lobes of the power pattern. It appears as a bright branch near the negative delay horizon delay limit. This feature is strongest at short antenna spacings and fades with increasing antenna spacing. The bright signature is absent in the {\it zenith} pointing (bottom left panel, Figure~\ref{fig:noiseless-dsm-csm-delay-spectrum}) because the Galactic center has set below the horizon. 

\subsubsection{Ubiquitous Diffuse Emission}\label{sec:diffuse-features}

Diffuse emission outside the Galactic plane manifests in the primary field of view as a branch at $\tau>0$ and $\tau=0$ in the {\it off-zenith} and {\it zenith} pointings respectively. The former is seen at $\tau>0$ because the primary lobe of the power pattern is centered eastward of zenith, whereas in the latter it is centered at zenith. As we see from Equation \ref{eqn:obsvis}, each baseline measures a single spatial mode on the sky with an angular size scale inversely proportional to the length of the baseline projected in the direction of the emission. Thus, in the {\it zenith} pointing, the horizontal line at $\tau=0$, fades away on antenna spacings $|\boldsymbol{b}| \gtrsim 125$~m because the diffuse sky model is devoid of spatial structures on scales $\lesssim$~0\fdg 75. 

\subsubsection{Diffuse Emission on Wide Antenna Spacings}\label{sec:diffuse-long-baselines}

In both pointings the diffuse emission (left panels) is prominent near the horizon delay limits extending to the widest antenna spacings. This is a characteristic signature of the wide-field effects discussed in \S\ref{sec:wide-field}. It is evident at all LSTs in our simulations. Thus, diffuse emission from far off-axis directions manifests as an edge-heavy {\it two-pronged fork} across all baselines. It decreases in strength with increasing baseline length but is nevertheless present in all baseline orientations.   

\subsubsection{Compact Foreground Signatures}\label{sec:compact}

In contrast to the delay power spectra of diffuse emission, compact emission (right panels) manifests as a center-heavy structure in either pointing. 

The amplitude response of an interferometer to a point source is, to first order, flat across baseline length. Since the primary field of view in the {\it off-zenith} pointing is centered eastward of zenith, the bulk of the compact foreground emission is seen in a branch with $\tau>0$. In the {\it zenith} pointing, compact emission from the same patch of sky is seen as a bright horizontal arm at $\tau=0$ since the primary field of view is centered at zenith. 

Foreground emission at $\tau=0$ and $\tau<0$ in the {\it off-zenith} pointing is caused by point sources co-located with secondary lobes of the power pattern. On the other hand, point sources co-located with secondary lobes of power pattern in the {\it zenith} pointing are revealed as faint but distinct branches at positive and negative delays depending on the orientation of antenna spacing and direction of emission on the sky. 

\subsection{The ``Pitchfork''}\label{sec:composite}

Delay spectra from the foreground model in our study display a composite feature set drawn from the features of compact and diffuse foreground models. Here we compare the relative strengths of emission from different spatial scales in our composite foreground model. 

When not dominated by the bright emission from the Galactic center, the delay power spectrum of the combined foreground model is composed of diffuse and compact emission both of which are significant. This is illustrated by a detailed examination of the {\it zenith} pointing. 

\begin{figure}[htb]
\centering
\includegraphics[width=\linewidth]{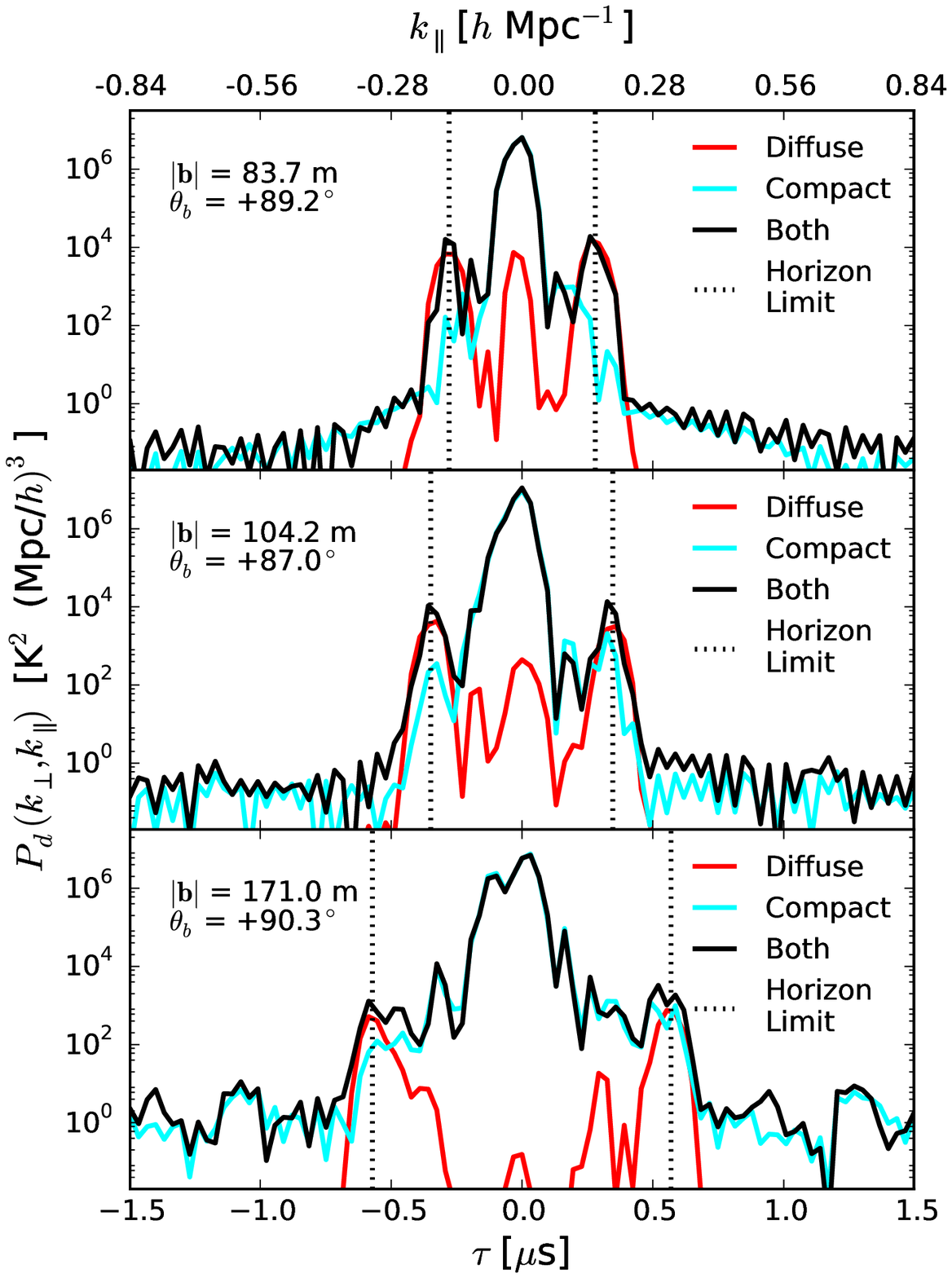}
\caption{Simulated delay power spectra for three chosen northward oriented antenna spacings of length: $\sim$84~m ({\it top}), $\sim$104~m ({\it middle}), and $\sim$171~m ({\it bottom}). The baseline length and orientation is specified in each panel. The solid red, cyan, and black lines denote contributions from diffuse, compact, and composite foreground models respectively. Vertical dotted lines mark the horizon delay limits. Compact emission dominates the central regions of the delay power spectra while both components, especially diffuse emission on short antenna spacings, dominate near the horizon delay limits, giving rise to a characteristic {\it pitchfork}-shaped structure. \label{fig:pitchfork-baselines}}
\end{figure}

Figure~\ref{fig:pitchfork-baselines} shows delay power spectra of three antenna pairs of different spacings oriented northward during the {\it zenith} pointing; each is a different vertical slice of the delay power spectra plots shown in Figure~\ref{fig:noiseless-dsm-csm-delay-spectrum}. The diffuse, compact, and composite components are shown as solid red, cyan, and black lines, respectively. The horizon delay limits are shown as a pair of vertical dotted lines. 

The peak at $\tau=0$ (corresponding to the primary lobe in the power pattern) with a value of $\sim 10^7$--$10^8$~K$^2$(Mpc/$h$)$^3$, independent of antenna spacing, is predominantly determined by compact emission. The peak at $\tau=0$ from diffuse emission is $\sim 10^3$ times fainter and decreases rapidly with increase in antenna spacing. This is the response expected from different antenna spacings toward compact and diffuse emission. 

Near the horizon delay limits, the diffuse component is brighter relative to the compact component. Here, diffuse emission does not decrease as rapidly with increasing antenna spacing as was seen at $\tau=0$. In fact, even on widely spaced antennas, diffuse emission in the delay power spectrum near the horizon delay limits exceeds that in the primary lobe by about three orders of magnitude. This feature is described in \S\ref{sec:diffuse-long-baselines} and attributed to wide-field measurement effects discussed in \S\ref{sec:wide-field}. 

Simulations with the complete foreground model show the combination of center-heavy features dominated by compact emission in primary field of view, and edge-heavy features from both types of emission especially the diffuse component near the horizon. This results in a characteristic {\it pitchfork} structure imprinted in the {\it foreground} wedge and should be evident in observations.

The observability of the {\it pitchfork} signature predicted in this paper depends on the relative levels of uncertainty in the foreground model and fluctuations from thermal noise. In our simulations, since thermal noise in these very short duration snapshots is $\sim 10^4$~K$^2$(Mpc/$h$)$^3$ and features near the horizon delay limits are also of comparable amplitudes, the {\it pitchfork} feature is not expected to be detected, though this feature is marginally visible in the {\it zenith} pointing of observed data (see Figure~\ref{fig:fhd-sim-comparison}). We attribute this to differences between our foreground model and the actual sky.  Deeper observations should reveal the feature clearly.

We also note that increasing the antenna spacing progressively improves the resolution along the delay axis by increasing the number of delay bins inside the {\it foreground wedge}. This improves the localization of foreground objects whose signatures are imprinted in the delay power spectrum. For instance, there is an increase in the number of secondary peaks in the delay power spectrum between $\tau=0$ and horizon delay limits as the antenna spacing increases from $\sim$84~m to $\sim$171~m. In this case, these correspond to secondary lobes of the power pattern that lie between the primary lobe and the horizon along the local meridian. At short antenna spacings, due to relatively lower resolution along the delay axis inside the {\it foreground wedge} and a consequent loss of localization of foreground emission, these secondary peaks blend in with other major peaks and are not distinctly visible. 

\section{Baseline-Based Foreground Mitigation}\label{sec:fg-grading}

Here, we investigate the susceptibility of particular antenna spacings to foreground contamination arising out of bright foreground objects located near the horizon and present a technique to substantially mitigate such contamination. We use the MWA as an example.

\begin{figure*}[htb]
\centering
\subfloat[][Sky model]{\label{fig:sky-model}\includegraphics[width=0.33\linewidth]{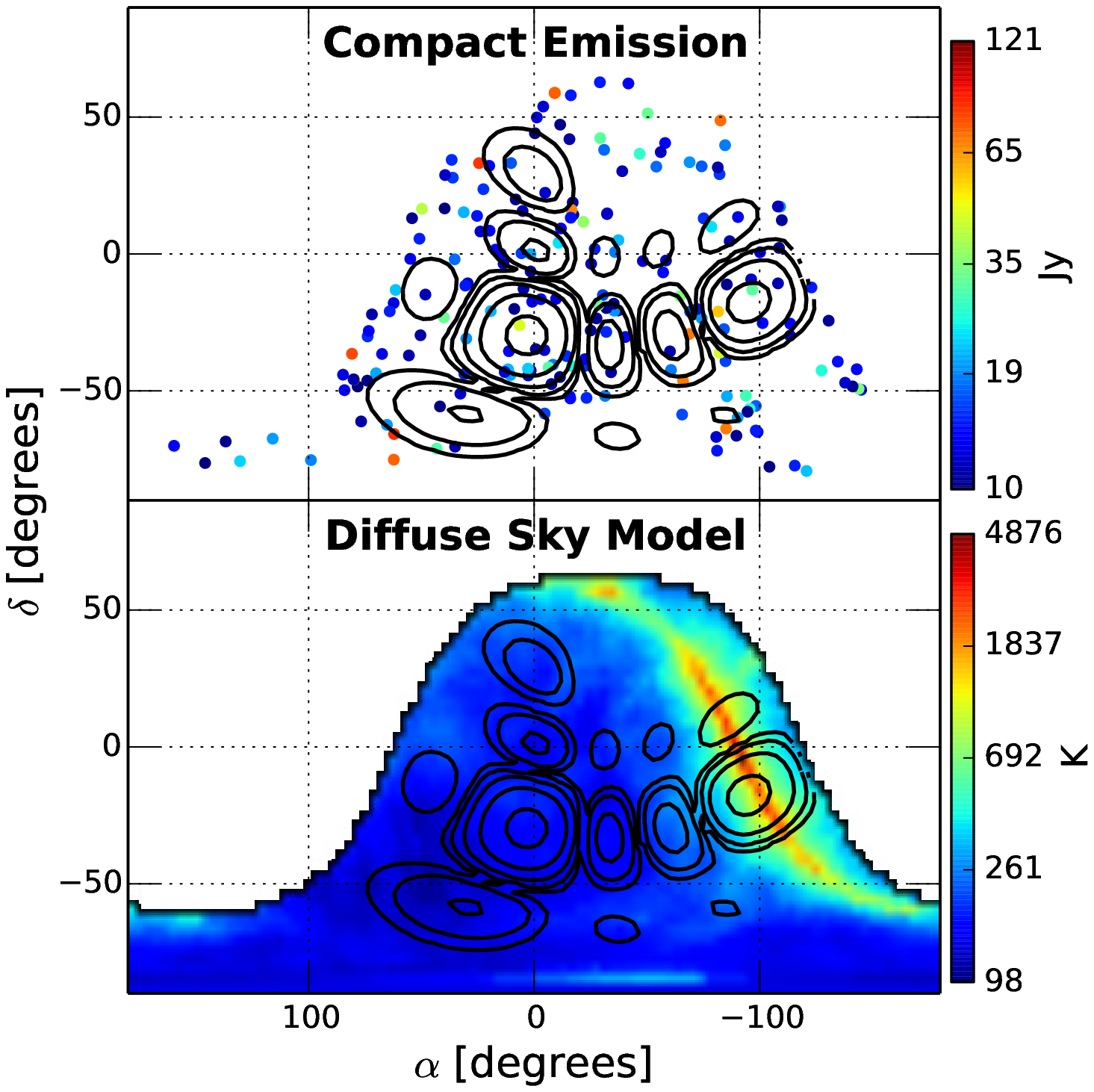}}
\subfloat[][Delay maps]{\label{fig:delay-map}\includegraphics[width=0.33\linewidth]{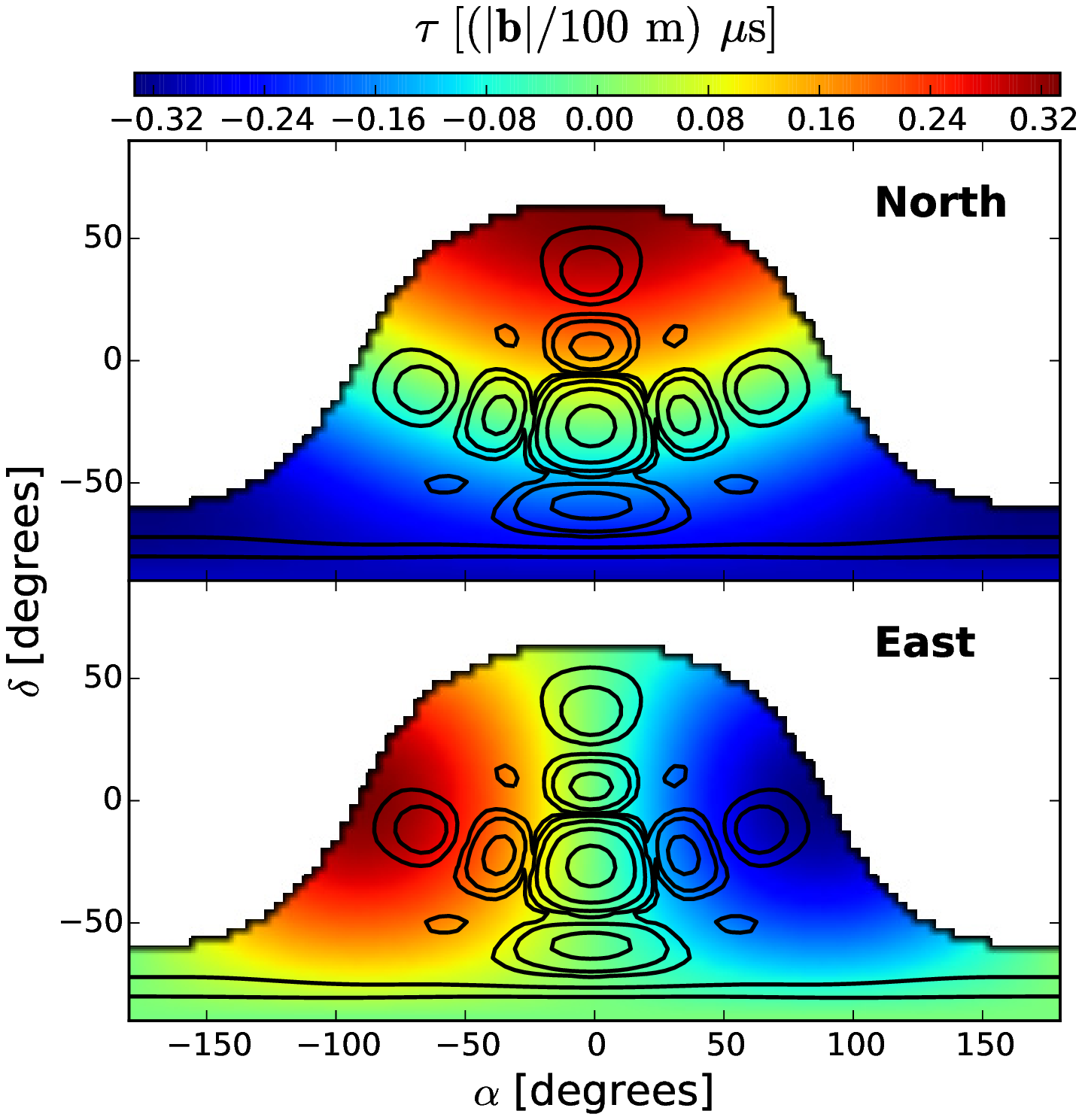}}
\subfloat[][Delay Spectra]{\label{fig:baseline-breakup}\includegraphics[width=0.33\linewidth]{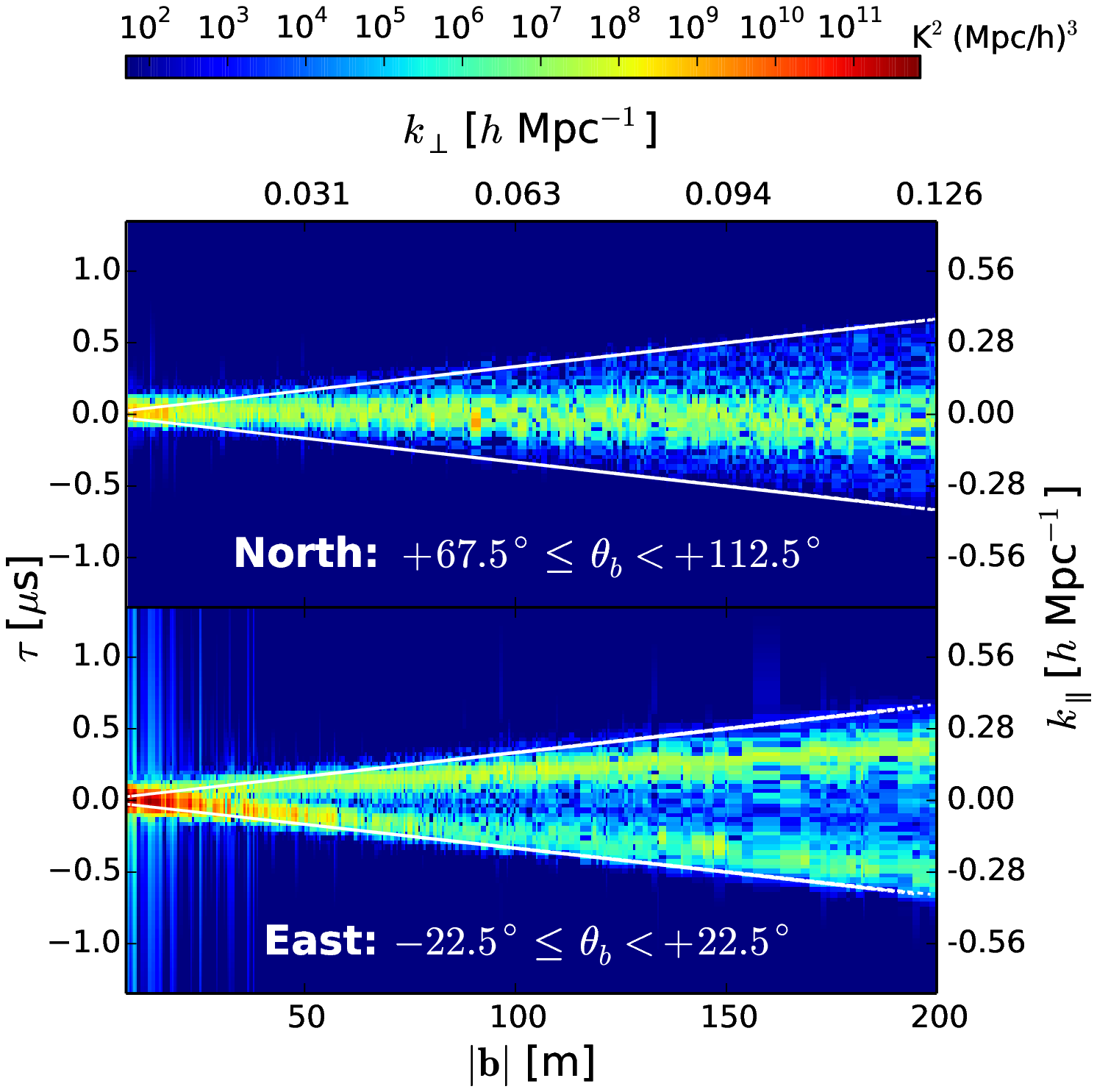}}
\caption{(a) Sky model showing compact ({\it top}) and diffuse ({\it bottom}) emission (adopted from Figure~\ref{fig:sky-model-MWA}). The Galactic center is very prominent in diffuse emission on the west co-located with a bright secondary lobe of the power pattern. (b) Sky hemisphere mapped to delays observed on antenna spacings oriented north ({\it top}) and east ({\it bottom}). Delays vary linearly with antenna spacing length. Color scale shown is for a 100~m antenna spacing. The bright Galactic center will appear at $\tau=0$ in northward antenna spacings and close to negative horizon delay limit on eastward antenna spacings. (c) Simulated delay power spectra on antenna spacings oriented northward ({\it top}) and eastward ({\it bottom}). White lines denote horizon delay limits. The bright Galactic center is prominently visible close to negative horizon delay limit, especially on short eastward antenna spacings. These are also the most severely contaminated by foreground spillover. The northward antenna spacings, on the other hand, are the least contaminated.}
\label{fig:breakup}
\end{figure*}

The Galactic center in the {\it off-zenith} pointing is one such example already available in our study. Figure~\ref{fig:sky-model} shows the sky model ({\it top:} compact component, {\it bottom:} diffuse component) in this pointing. The Galactic center is the most dominant source of foreground contamination from the diffuse sky model and is co-located with a bright secondary lobe of the power pattern near the western horizon. Figure~\ref{fig:delay-map} shows the sky mapped to delays registered by the baseline vectors, of length 100~m for instance, oriented toward north (top panel) and east (bottom panel). Figure~\ref{fig:baseline-breakup} shows the delay spectra on baselines oriented northward ($67\fdg 5\le \theta_\textrm{b} < 112\fdg 5$) at the top and eastward ($-22\fdg 5~\le \theta_\textrm{b} < 22\fdg 5$) at the bottom. The Galactic center manifests itself most distinctly near the negative horizon delay limit on short eastward baselines in the delay power spectrum (bottom panel of Figure~\ref{fig:baseline-breakup}). Consequently, the spillover caused by the instrument's spectral transfer function from the {\it foreground wedge} into the {\it EoR window} affects the northward baselines the least and is most severe on eastward baselines (particularly the short ones) evident by the bright vertical stripes of foreground contamination. 

With a foreground model known {\it a priori} in which structures and locations of very bright foreground objects such as the Galactic center or AGN are available, we can predict the response across antenna spacings as a function of observing parameters such as LST, power pattern, etc. This allows us to programmatically screen data for antenna spacings that are severely contaminated by foregrounds near the horizon delay limits. These can be weighted appropriately during data analysis. We demonstrate such a screening technique, whereby we use the bright object's location and structure to discard antenna spacings of certain lengths and orientations to mitigate foreground contamination in the {\it EoR window}. 

\begin{figure}[htb]
\centering
\includegraphics[width=\linewidth]{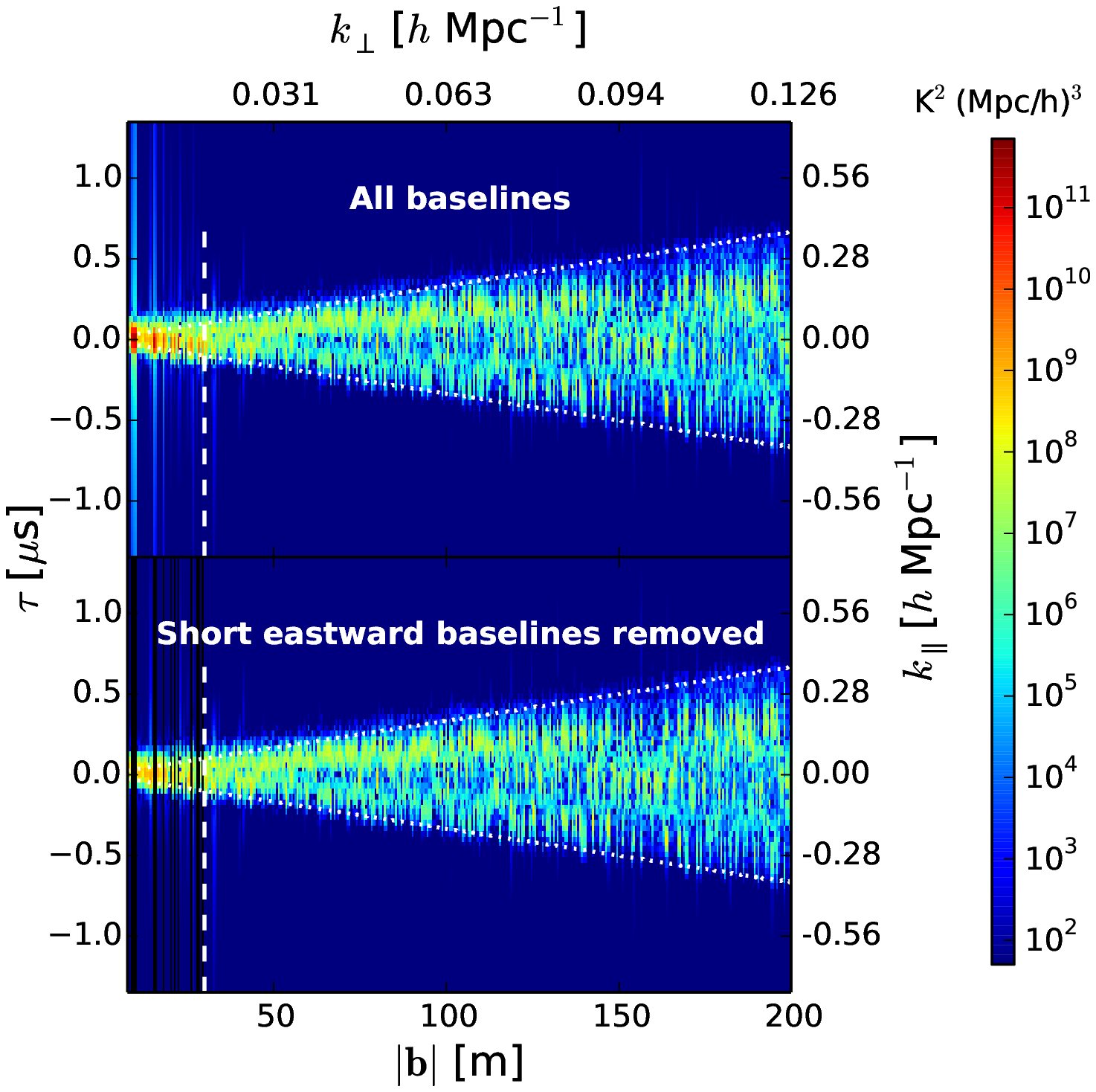}
\caption{Simulated delay spectra power for the {\it off-zenith} pointing with all antenna spacings included ({\it top}) and with short eastward antenna spacings discarded ({\it bottom}). Discarded antenna spacings (black vertical stripes) have lengths $|\boldsymbol{b}|<30$~m (leftward of vertical dashed line) and orientations $|\theta_\textrm{b}|<$~22\fdg 5. The spillover from the bright Galactic center near the negative horizon delay limit from the {\it foreground wedge} is lowered by a factor $\sim 100$ when short eastward antenna spacings are discarded. \label{fig:before-after}}
\end{figure}

In our example, we discard eastward antenna spacings ($-22\fdg 5\le\theta_\textrm{b}< 22\fdg 5$) of lengths $|\boldsymbol{b}| < 30$~m. Foreground contamination was found to be insensitive to removal of wider antenna spacings as discussed below. Figure~\ref{fig:before-after} shows the delay spectra obtained with all antenna spacings (top panel) and after applying our screening technique (bottom panel) on the {\it off-zenith} observation. Notice the significant reduction in foreground spillover into the {\it EoR window} via the removal of bright vertical stripes on short eastward antenna spacings. 

\begin{figure}[htb]
\centering
\includegraphics[width=\linewidth]{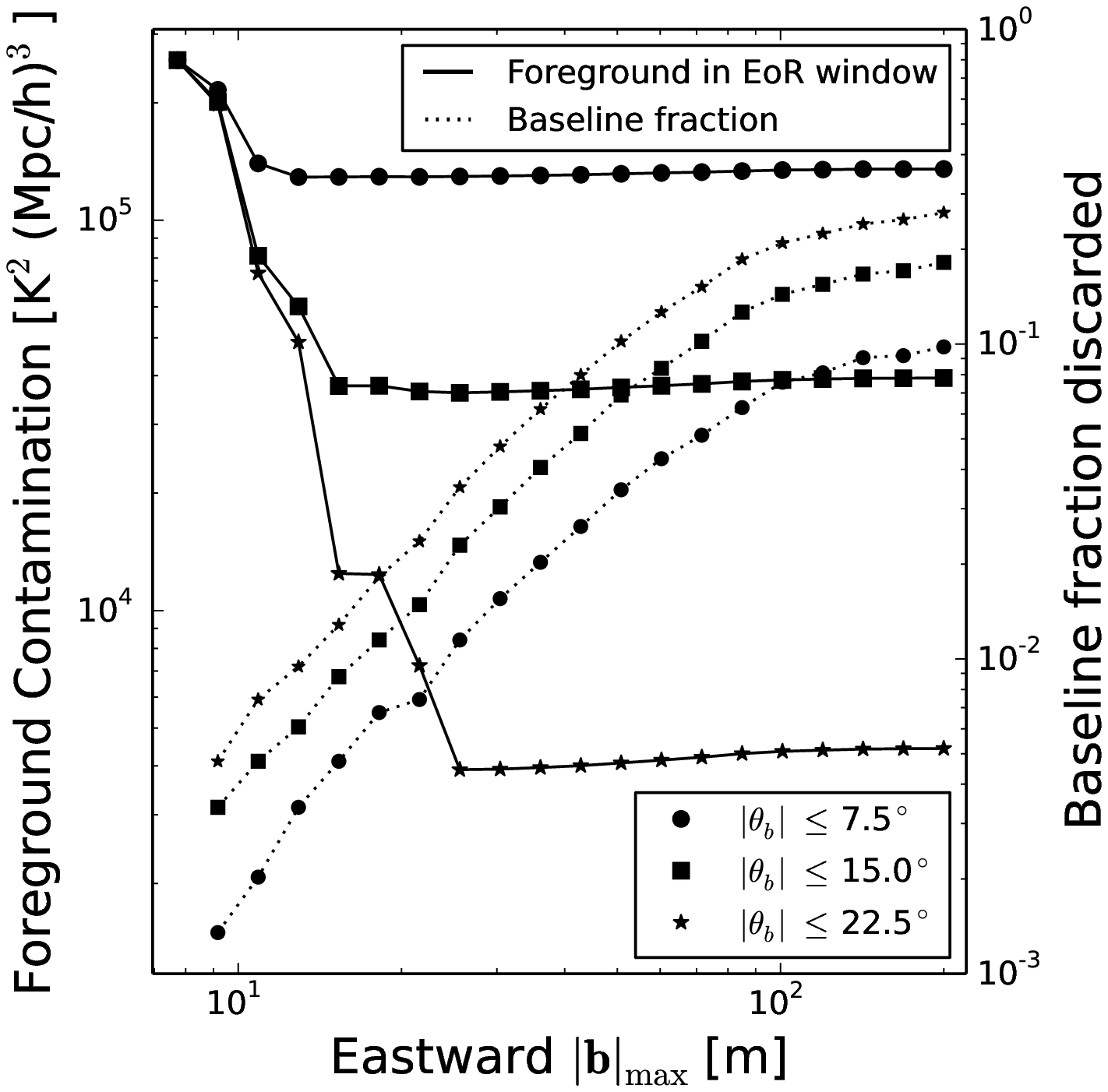}
\caption{Drop in foreground contamination in the MWA {\it EoR window}, and loss of data for the {\it off-zenith} pointing as a function of discarded baselines. Eastward baselines with varying degrees of directedness --- $|\theta_\textrm{b}|<$~7\fdg 5 (solid circles), $|\theta_\textrm{b}|<$~15\arcdeg~ (solid squares), and $|\theta_\textrm{b}|<$~22\fdg 5 (solid stars) --- and lengths $|\boldsymbol{b}| \le |\boldsymbol{b}|_\textrm{max}$ ($x$-axis) are discarded. Loss of data (dotted lines) is measured by discarded baselines as a fraction of the total number for the corresponding cases. Foreground contamination in the {\it EoR window} (solid lines) drops by a factor $\sim$2 ($|\theta_\textrm{b}|\le$~7\fdg 5) to $\sim$100 ($|\theta_\textrm{b}|\le$~22\fdg 5). The latter limit can be achieved with a mere 5\% loss of data at $|\boldsymbol{b}|_\textrm{max}\simeq30$~m, and discarding longer baselines ($|\boldsymbol{b}|\gtrsim 30$~m) has no effect in further reducing foreground contamination. \label{fig:screening}}
\end{figure}

This screening technique can be generalized to optimize between foreground mitigation and loss of sensitivity from discarding data. Figure~\ref{fig:screening} shows how the typical foreground contamination\footnote{Foreground contamination is measured by standard deviation of noiseless $P_\textrm{d}(\boldsymbol{k}_\perp,k_\parallel)$ from foregrounds in the MWA {\it EoR window}.} in the MWA {\it EoR window} depends on the orientations and lengths of discarded antenna spacings. We choose antenna spacings oriented eastward to varying degrees of directedness, i.e., $-7\fdg 5~\le\theta_\textrm{b} < 7\fdg 5$ (solid circles), $-15\arcdeg\le\theta_\textrm{b}< 15\arcdeg$ (solid squares), and $-22\fdg 5\le\theta_\textrm{b}< 22\fdg 5$ (solid stars). Among antenna spacings that satisfy these criteria, we discard data from those whose lengths are shorter than $|\boldsymbol{b}|_\textrm{max}$ ($x$-axis) and show foreground contamination estimated in the {\it EoR window} from all remaining antenna spacings. 

In other words, Figure~\ref{fig:screening} demonstrates the progress in foreground mitigation as orientation and maximum length of discarded antenna spacings are varied. The fraction of discarded antenna spacings discarded relative to the total number is shown in dotted lines for different ranges of $\theta_\textrm{b}$. It is seen that foreground contamination can be mitigated by a factor between $\sim 2$ ($|\theta_\textrm{b}|\le$~7\fdg 5) and $\sim 100$ ($|\theta_\textrm{b}|\le$~22\fdg 5). The latter limit is achieved with a mere 5\% loss of data for $|\boldsymbol{b}|_\textrm{max}\simeq30$~m. Discarding antenna spacings with lengths $|\boldsymbol{b}|\gtrsim 30$~m does not mitigate foreground contamination any further and would only lead to loss of sensitivity as the fraction of discarded baselines increases from $\sim 5$\% to $\sim 25$\%. 

If there was a bright point source in place of the Galactic center, it will give rise to foreground contamination even on longer antenna spacings. Such cases will necessitate discarding more or all of the eastward antenna spacings. The MWA was used as an example. In general, the direction, strength and type of emission, and the array layout will determine such thresholds.

In principle, instead of discarding selected antenna spacings altogether, we could down-weight them based on an optimal scheme. For instance, the estimates of covariance computed from the delay transform bins can be naturally fed into the covariance-weighted power spectrum estimation techniques \citep{liu14a,liu14b}. It could also be used to downweight or flag contaminated baselines in imaging applications. This technique provides a very simple and yet effective tool to suppress the effects of foreground contamination in EoR data analysis. 

\section{Summary}\label{sec:summary}

Our primary motivation in this work is to understand how the various bright foregrounds will manifest in three-dimensional power spectrum of H~{\sc i} from 21~cm reionization observations. In units of temperature variance, the dynamic range between bright foregrounds and the 21~cm signal is expected to be $\sim 10^8$; a detailed understanding of how foregrounds can corrupt the 21~cm power spectrum is therefore essential. This analysis extends previous work by simulating the entire sky rather than just the central field of view and by providing a comparison with early observations with the MWA. By making use of the delay spectrum technique to estimate the power spectrum, we are able to observe the effects of foregrounds while avoiding entanglements with more sophisticated power spectrum estimators.  

We find that all wide-field instruments, typical of modern EoR observatories, imprint a characteristic {\it two-pronged fork} signature in delay spectra. This arises from two related effects: delay bins near the horizon subtend larger solid angles and therefore contain larger integrated emission; and, foreshortening of baselines toward the horizon makes them sensitive to emission on large angular scales which match the inverse of their foreshortened lengths. These effects combined with higher sensitivity of antennas in the primary field of view results in a characteristic {\it pitchfork} signature. The amplitude of these generic signatures can be controlled by careful design of antenna aperture. In contrast to a dipole and a phased array such as an MWA tile, a dish such as the one proposed for HERA is found to yield the least foreground contamination and thus preferable for EoR studies.

Simulating in many important respects the response of the MWA to an all-sky foreground model that consists of diffuse Galactic emission and bright point sources, we confirm that the modeled delay spectra are in agreement with data obtained with the MWA to within expected uncertainties in foreground models. 

Our simulations enable us to identify numerous signatures of different components of foreground emission seen in the delay spectra. We establish the relationship between these signatures and observing parameters such as antenna pointing and LST, instrument parameters such as antenna power pattern, and foreground parameters such as type of emission, etc. 

The bright Galactic center at the edge of the western horizon co-located with one of the far secondary lobes of MWA tile power pattern is the brightest source of foreground contamination in the {\it off-zenith} pointing. It manifests itself near the negative horizon delay limit in the delay power spectrum on eastward antenna spacings. 

Diffuse emission in the primary field of view is prominent on shorter antenna spacings. However, it is also prominent near the horizon limits even on wide antenna spacings --- an effect of the wide-field nature of the measurement. On the other hand, compact emission predominantly maps onto central regions of the {\it foreground wedge}. Features arising from compact emission co-located with primary and secondary lobes of the antenna power pattern have been identified. In general, delay power spectrum signatures of compact emission are center-heavy while those of diffuse emission are edge-heavy which results in a characteristic {\it pitchfork} signature. This will be evident when thermal noise is sufficiently lowered, as larger volumes of data are processed. 

We also provide a simple and effective tool based on the delay spectrum technique that can potentially mitigate foreground contamination by nearly two orders of magnitude in EoR data analysis by discarding or down-weighting data from antenna pairs most affected by foreground contamination, with negligible loss of sensitivity. In conclusion, we find that inclusion of emission models, both diffuse and compact, all the way to the horizon is essential to explaining the observed power spectrum. 

\acknowledgments

This work was supported by the U. S. National Science Foundation (NSF) through award AST-1109257. DCJ is supported by an NSF Astronomy and Astrophysics Postdoctoral Fellowship under award AST-1401708. JCP is supported by an NSF Astronomy and Astrophysics Fellowship under award AST-1302774. This work makes use of the Murchison Radio-astronomy Observatory, operated by CSIRO. We acknowledge the Wajarri Yamatji people as the traditional owners of the Observatory site. Support for the MWA comes from the NSF (awards: AST-0457585, PHY-0835713, CAREER-0847753, and AST-0908884), the Australian Research Council (LIEF grants LE0775621 and LE0882938), the U.S. Air Force Office of Scientific Research (grant FA9550-0510247), and the Centre for All-sky Astrophysics (an Australian Research Council Centre of Excellence funded by grant CE110001020). Support is also provided by the Smithsonian Astrophysical Observatory, the MIT School of Science, the Raman Research Institute, the Australian National University, and the Victoria University of Wellington (via grant MED-E1799 from the New Zealand Ministry of Economic Development and an IBM Shared University Research Grant). The Australian Federal government provides additional support via the Commonwealth Scientific and Industrial Research Organisation (CSIRO), National Collaborative Research Infrastructure Strategy, Education Investment Fund, and the Australia India Strategic Research Fund, and Astronomy Australia Limited, under contract to Curtin University. We acknowledge the iVEC Petabyte Data Store, the Initiative in Innovative Computing and the CUDA Center for Excellence sponsored by NVIDIA at Harvard University, and the International Centre for Radio Astronomy Research (ICRAR), a Joint Venture of Curtin University and The University of Western Australia, funded by the Western Australian State government.  

\appendix

\section{Delay Transform Conventions}\label{sec:delay-conventions}

\begin{figure}[htb]
\centering
\includegraphics[width=\linewidth]{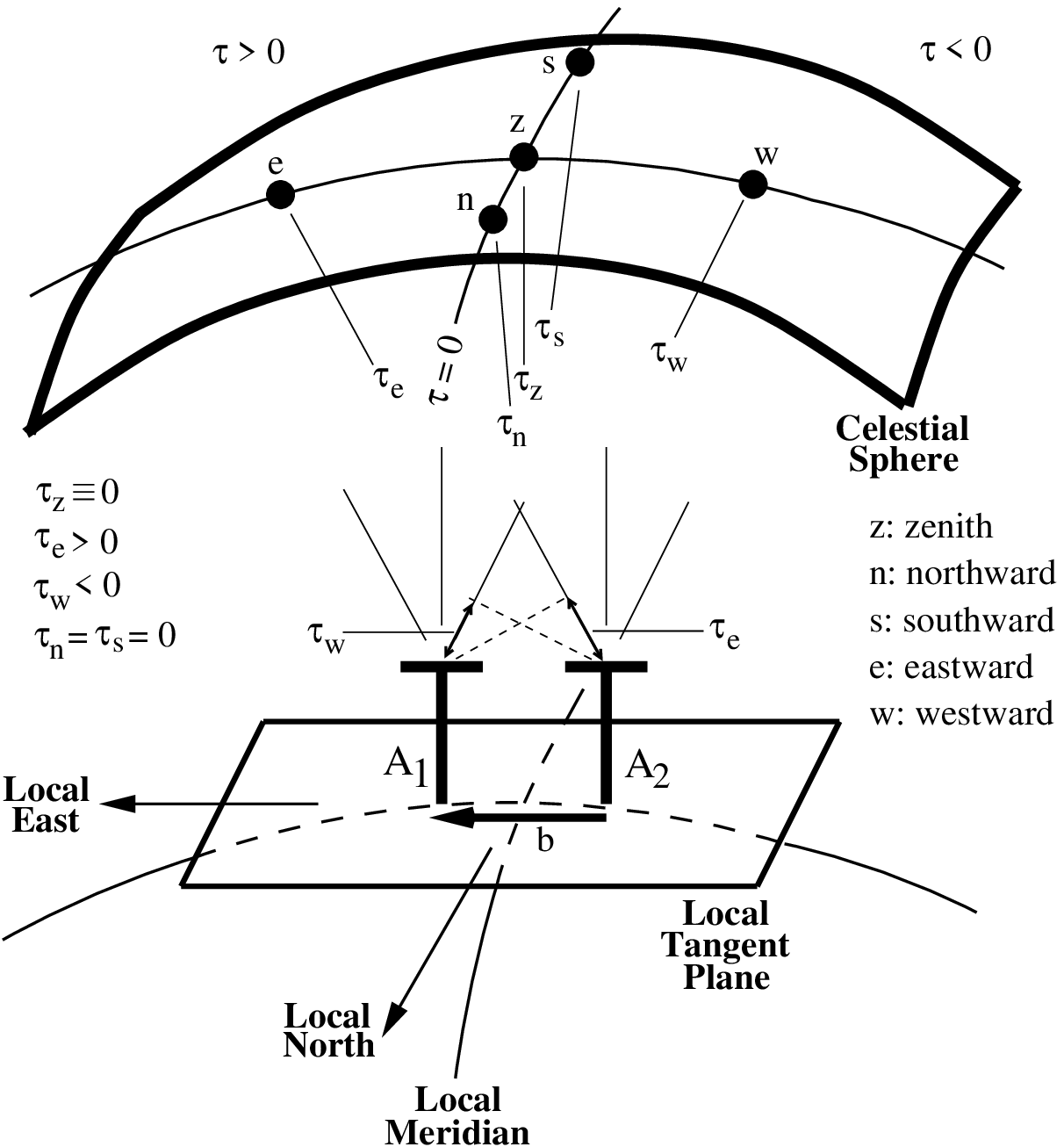}
\caption{Radio interferometer delay conventions used in this paper. Two antennas labeled as A$_1$ and A$_2$ are separated by vector $\boldsymbol{b}$ on the local tangent plane. The local meridian, local north and local east are shown for reference. Points labeled as `z', `n', `s', `e' and `w' on the celestial sphere denote zenith, northward, southward, eastward and westward positions, respectively. $\tau_\textrm{z}$, $\tau_\textrm{n}$, $\tau_\textrm{s}$, $\tau_\textrm{e}$ and $\tau_\textrm{w}$ denote the respective delays measured between A$_1$ and A$_2$. Throughout this paper, zenith is chosen as the phase center. Hence, $\tau_\textrm{z}\equiv 0$. If $\boldsymbol{b}$ is oriented eastward as shown, then $\tau_\textrm{e} > 0$, $\tau_\textrm{w} < 0$, and $\tau_\textrm{n}=\tau_\textrm{s}=0$. Conversely, if $\boldsymbol{b}$ is oriented northward (not shown here), then $\tau_\textrm{n} > 0$, $\tau_\textrm{s} < 0$, and $\tau_\textrm{e}=\tau_\textrm{w}=0$.} \label{fig:delay-cartoon}
\end{figure}

Figure~\ref{fig:delay-cartoon} illustrates the radio interferometer delays and conventions used in our paper. $\boldsymbol{b}$ is assumed to be on a coordinate system aligned with the local east, north (along local meridian) and upward (zenith) directions at the telescope site. Hence, a perfectly eastward oriented antenna spacing will observe objects in the eastern and the western skies at $\tau>0$ and $\tau<0$, respectively. Similarly, an object in the northern sky will appear at $\tau>0$ for an antenna spacing oriented northward. For all observations used in this study, we use zenith as the phase center, for which $\tau\equiv 0$. 

\section{Cosmological H~{\sc i} Power Spectrum}\label{sec:EoR-power-spectrum}

Equation~\ref{eqn:obsvis} can be equivalently expressed as:
\begin{align}\label{eqn:vis}
  V_u(f) &= \iint\limits_\textrm{sky} A(\hat{\boldsymbol{s}},f)\,I(\hat{\boldsymbol{s}},f)\,W_\textrm{i}(f)\,e^{-i2\pi\boldsymbol{u}\cdot\hat{\boldsymbol{s}}}\,\dif\Omega,
\end{align}
where, $\hat{\boldsymbol{s}}$ is measured with reference to a location on the sky referred to as the {\it phase center}, and $\boldsymbol{u}\equiv (u,v,w)$ denotes the spatial frequency vector. $w$ is aligned parallel to the direction of the phase center, while $u$ and $v$ lie on the transverse plane perpendicular to it. For measurements that lie on this plane, we can choose $w=0$ without loss of generality and $\boldsymbol{u}$ effectively reduces to $\boldsymbol{u}\equiv (u,v)$, a two-dimensional vector. Then, $\boldsymbol{u}$ is directly related to the transverse spatial wavenumber mode as: 
\begin{align}\label{eqn:k-perp}
  \boldsymbol{k}_\perp &\equiv \frac{2\pi\boldsymbol{u}}{D}, 
\end{align}
where, $D\equiv D(z)$ is the transverse comoving distance at redshift $z$. 

Since we are concerned with a redshifted H~{\sc i} spectral line from cosmological distances, $f$ is a measure of cosmological distance along the line of sight. $\eta$, which is the Fourier transform dual of $f$, is used to denote the spatial frequency along the line of sight and has units of time. It is directly related to the line of sight wavenumber, 
\begin{align}\label{eqn:k-prll}
  k_\parallel &\approx \frac{2\pi\eta\,f_{21}H_0\,E(z)}{c(1+z)^2}, 
\end{align}
where, $f_{21}$ is the rest frame frequency of the 21~cm spin flip transition of H~{\sc i}, and $H_0$ and $E(z)\equiv [\Omega_\textrm{M}(1+z)^3+\Omega_\textrm{k}(1+z)^2+\Omega_\Lambda]^{1/2}$ are standard terms in cosmology. This approximation holds under the assumption that the redshift range (or frequency band) is small enough within which cosmological evolution is negligible. Thus,
\begin{align}\label{eqn:los-transform}
  \tilde{V}_u(\eta) &\equiv \int V_u(f)\,W(f)\,e^{i2\pi f\eta}\,\dif f
\end{align}
represents the true spatial Fourier representation of the three-dimensional sky brightness distribution. This approach has been discussed in detail in \citet{mor04}. The spatial power spectrum of EoR H~{\sc i} distribution, $P(\boldsymbol{k}_\perp,k_\parallel)$, and $\tilde{V}_u(\eta)$ are related by \citep{mor04,mcq06,par12a}: 
\begin{align}\label{eqn:true-power_spectrum}
  P(\boldsymbol{k}_\perp,k_\parallel) &\simeq |\tilde{V}_u(\eta)|^2\left(\frac{A_\textrm{e}}{\lambda^2\Delta B}\right)\left(\frac{D^2\Delta D}{\Delta B}\right)\left(\frac{\lambda^2}{2k_\textrm{B}}\right)^2,
\end{align}
where, $A_\textrm{e}$ is the effective area of the antenna, $\Delta B$ is the bandwidth, $\Delta D$ is the comoving depth along the line of sight corresponding to $\Delta B$, $\lambda$ is the wavelength of the band center, and $k_\textrm{B}$ is the Boltzmann constant. Thus, $\tilde{V}_u(\eta)$ inferred from observations, in units of Jy~Hz, can be converted into an equivalent cosmological H~{\sc i} power spectrum $P(\boldsymbol{k}_\perp,k_\parallel)$, in units of K$^2$~Mpc$^3$ or, more generally, K$^2$(Mpc/$h$)$^3$, where $h$ is the Hubble constant factor.

Without loss of generality, the phase center can be assumed to be the zenith relative to the local tangent plane. Then $\boldsymbol{u}$ lies on this plane for measurements constrained to be on it. If the array of antennas are also coplanar lying on the local tangent plane, then $\boldsymbol{u}=\boldsymbol{b}/\lambda$. Under such circumstances, equations~\ref{eqn:obsvis} and \ref{eqn:delay-transform} closely resemble equations~\ref{eqn:vis} and \ref{eqn:los-transform} respectively. However, they are not quite identical to each other. It is because $\boldsymbol{b}$ is independent of frequency while $\boldsymbol{u}$ is not. \citet{par12b} and \citet{liu14a} have discussed the mathematical correspondence between the two. 

\bibliographystyle{apj}
\bibliography{eor}

\end{document}